\documentclass[a4paper,12pt]{article}
\pdfoutput=1 
\usepackage{jcappub} 
\def\lsim{\raise0.3ex\hbox{$\;<$\kern-0.75em\raise-1.1ex\hbox{$\sim\;$}}}
\def\gsim{\raise0.3ex\hbox{$\;>$\kern-0.75em\raise-1.1ex\hbox{$\sim\;$}}}

\usepackage{hyperref}
\usepackage{amsmath}
\usepackage{relsize}
\usepackage{psfrag}
\usepackage{epsfig}
\usepackage{xcolor}
\usepackage{color}
\usepackage{graphics}
\usepackage{url}
\usepackage{relsize}
\usepackage[T1]{fontenc} 
\usepackage[percent]{overpic}
\usepackage{subcaption}
\usepackage{bigints}

 \normalsize

\graphicspath{{./figures/}}
\newcommand{\n}{\nonumber\\}

\begin{document}
\begin{flushright}
\end{flushright}
\title{Modified hybrid inflation, reheating and stabilization of the electroweak vacuum}
\author[a,1]{Merna~Ibrahim,}
\author[b,2]{Mustafa~Ashry,}
\author[a,3]{Esraa~Elkhateeb}
\author[a,4]{Adel~M.~Awad}
\author[c,5]{and Ahmad~Moursy}

\affiliation[a]{Department of Physics, Faculty of Science, Ain Shams University, 11566, Cairo, Egypt.}
\affiliation[b]{Department of Mathematics, Faculty of Science, Cairo University, 12613, Giza, Egypt.}
\affiliation[c]{Department of Basic Sciences, Faculty of Computers and Artificial Intelligence, Cairo University, Giza 12613, Egypt.}
\emailAdd{mernaibrahim\_p@sci.asu.edu.eg}
\emailAdd{mustafa@sci.cu.edu.eg}
\emailAdd{dr.esraali@sci.asu.edu.eg}
\emailAdd{a.awad@sci.asu.edu.eg}
\emailAdd{a.moursy@fci-cu.edu.eg}
\date{\today}
\abstract{We propose a modification to the standard hybrid inflation model \cite{Linde:1993cn}, that connects a successful hybrid inflation scenario to the standard model Higgs sector, via the electroweak vacuum stability. The proposed model results in an effective inflation potential of a hilltop-type, with both the trans-Planckian and sub-Planckian inflation regimes consistent with the recent Planck/BICEP combined results. Reheating via the inflation sector decays to right-handed neutrinos is considered. An upper bound on the reheating temperature $T_{\rm R} \lsim 2\times 10^{11}~(1\times 10^{13})$ GeV, for large~(small) field inflation, will suppress contributions from one-loop quantum corrections to the inflation potential. This may push the neutrino Yukawa couplings to be ${\cal O}(1)$ and affect the vacuum stability. 
 We show that the couplings of the SM Higgs to the inflation sector can guarantee the electroweak vacuum stability up to the Planck scale. The so-called hybrid Higgs-inflaton model leads to a positive correction for the Higgs quartic coupling at a threshold scale, which is shown to have a very significant effect in stabilizing the electroweak vacuum. We find that even with ${\cal O}(1)$ neutrino Yukawa couplings, threshold corrections leave the SM vacuum stability intact. 
}
\maketitle
\section{\label{sec:intro}INTRODUCTION}
The Standard Model of cosmology (SMC), and the Standard Model of particle physics (SM) are extremely successful in describing the observations of the Cosmic Microwave Background (CMB), and the low-energy experimental results of particle colliders, respectively. It turns out that elementary scalar particles play an important role in both particle physics and the early universe. 
In fact, the detection of the SM Higgs boson in 2012~\cite{ATLAS:2012yve,CMS:2012qbp} was the first successful signature of an elementary scalar playing a crucial role in particle physics. However, this discovery left unanswered questions, indicating that the SM is not the ultimate theory, such as the hierarchy problem and the problem of stability of the electroweak (EW) vacuum.

On the other hand, a scalar field (the inflaton $\phi$) is believed to play another pivotal role in the early universe, where it may be responsible for the cosmic inflation. The latter resolves the problems of flatness and horizon of the SMC, and the absence of early phase transition remnants can be justified. It is tempting to investigate possible connections between the two sectors of inflation and particle physics as well as impacts on both high- and low-energy physics. One portal to such a connection is the reheating process after the end of inflation, where the inflaton oscillates around its true minimum and decays into the SM particles when the Hubble parameter and the inflaton decay are of the same order, $H\sim \Gamma_\phi$. In nonoscillatory models \cite{BuenoSanchez:2007jxm,Dimopoulos:2017zvq,Heurtier:2022rhf}, where the inflaton keeps rolling in a runway direction, reheating after inflation is achieved via different mechanisms such as gravitational reheating, instant preheating, or curvaton reheating for example. Other connections between the two sectors can be achieved via a messenger that interacts with both sectors and influences physics in both of them as studied in~\cite{Masina:2011un,Masina:2011aa,Masina:2012yd,Saha:2016ozn,Lebedev:2021xey}.

The SM Higgs vacuum stability is one of the issues that raises concerns in both beyond standard model particle physics and cosmology of the early universe. The EW vacuum is stable up to an instability scale $\Lambda_I\sim 10^{11}$, where for higher scales the SM Higgs quartic coupling is driven to negative values by the dominant contribution of the top quark Yukawa coupling to the RGEs. As a matter of fact, the EW vacuum stability is very sensitive to the precision measurements of the top quark mass $m_t$ and less sensitive to the strong coupling $\alpha_s$~\cite{Elias-Miro:2011sqh,Datta:2013mta,Branchina:2014usa}. According to these uncertainties, the instability scale lies between $10^9 \lesssim \Lambda_I \lesssim 10^{12} $ GeV.
However, in the presence of a deeper minimum of the Higgs potential, at much larger field value~\cite{Coleman:1973jx,Sher:1988mj}, the EW vacuum may be metastable if its lifetime is larger than the age of the universe. The current measurements of the top quark mass $m_t$ and the Higgs mass $m_h$ support the hypothesis that the EW vacuum is metastable~\cite{Elias-Miro:2011sqh,Elias-Miro:2012eoi}.
 However, this situation may be subject to change for more precise measurements of $m_t$, and even considering new physics that explains the neutrino masses. Accordingly, the EW vacuum may move to the instability region. Moreover, quantum fluctuations may push the Higgs over its barrier causing destabilization of the Higgs during inflation~\cite{Gross:2015bea,Fumagalli:2019ohr,Bezrukov:2019ylq}. If typical momentum, which is of the same order as the hubble scale during inflation $k\sim H^{\rm inf}$, is greater than the potential barrier, then the EW vacuum can decay. Therefore, considering the inflation sector may even worsen the situation of the EW vacuum stability/metastability. These problems can be avoided if new physics arises at the instability scale $\Lambda_I$ or by defining a direct coupling between the inflaton and the Higgs boson~\cite{Lebedev:2012sy,Gross:2015bea,Ema:2017ckf,Lebedev:2021xey}.

The hybrid inflation model (HI)~\cite{Linde:1993cn} combines the inflation potential with a spontaneous symmetry breaking potential, where the inflation ends by a waterfall phase triggered by the inflaton $\phi$. In its simplest form, this class of models predicts a large spectral index $ns\sim 1$ and very small tensor-to-scalar ratio $r$ if the field variations are taken to be as small as sub-Planckian values~\cite{Linde:1993cn,Rehman:2009wv}. On the other hand, if field variations are super-Planckian, we have $r>0.1$. Both limits of the model are ruled out by Planck observations ~\cite{BICEP:2021xfz,Planck:2018jri}. In Ref.~\cite{Rehman:2009wv}, it was indicated that including one-loop quantum corrections to HI tree-level potential can improve the $n_s$ and $r$ values. It was assumed that the inflation scalars interact with right-handed neutrinos (RHN), that acquire large masses through the waterfall scalar vev. In this case, the neutrino Yukawa coupling with SM Higgs can be $\sim {\cal O}(1)$, which worsens the EW vacuum stability~\cite{Datta:2013mta} and may even be dangerous for the metastability. 

In this paper, we propose a connection between the SM Higgs sector and the hybrid inflation sector. We introduce a new scalar field $\chi$ that is interacting with the hybrid inflation sector to improve the inflation observables for a tree-level scalar potential on one hand, and on the other hand, stabilizes the EW vacuum up to the Planck scale. We will make use of the threshold modifications to the running of the SM Higgs coupling due to the inflation sector fields.

The paper is organized as follows. In Sec.~\ref{sec:model}, we provide the details of the modified hybrid inflation model as well as the inflation dynamics and the effective inflation potential. In Sec.~\ref{sec:observables}, we study the parameter space and the observables predictions of the model. Then we explore constraints from reheating, neutrino masses, and quantum corrections in Sec.~\ref{sec:reheat}. Section~\ref{sec:vs} is devoted to investigate the low-energy consequences such as the stability of the SM Higgs vacuum. Finally, we give our conclusions in Sec.~\ref{sec:conclusions}.

\section{\label{sec:model}MODIFIED HYBRID INFLATION MODEL} 
We propose a modified version of the hybrid inflation model~\cite{Linde:1993cn} (MHI). It consists of three SM singlet real scalars, the inflaton $\phi$ and the waterfall field $\psi$ as well as an extra scalar field $\chi$, with full scalar potential of the form
\begin{align}
V_{\rm MHI}(\phi,\psi,\chi)&= \lambda_\psi \Big(\psi^2 - \dfrac{v_{\psi}^2}{2} \Big)^2 + \frac{m^2}{2}\phi^2 + 2 \lambda_{\phi\psi} \phi^2 \psi^2 - 2\lambda_{\phi\chi} \phi^2 \chi^2 \n
&+\lambda_\chi \Big(\chi^2 - \frac{v_\chi^2}{2}\Big)^2 +2\lambda_{\psi\chi}\Big(\psi^2 - \dfrac{v_{\psi}^2}{2} \Big) \Big( \chi^2 - \frac{v_\chi^2}{2}\Big),
\label{eq:vmhi} 
\end{align}
where $m,~v_{\psi},~v_\chi$ are mass dimensionful scales while $\lambda_\psi,~\lambda_\chi,~\lambda_{\phi\psi},~\lambda_{\phi\chi},~\lambda_{\psi\chi}$ are dimensionless couplings. The first three terms correspond to the known hybrid inflation model~\cite{Linde:1993cn}. The last three terms give the modification on the HI\footnote{Similar modifications on chaotic inflation were also proposed in~\cite{Harigaya:2015pea,Saha:2016ozn}, based on a shift symmetry arguments.}.
The negative coefficient in the fourth term of (\ref{eq:vmhi}) can be justified in the context of the inverted hybrid inflation (IHI)~\cite{Lyth:1996kt} and may be obtained in some contexts such as supersymmetry~\cite{Lyth:1996kt}. \footnote{ In fact, we do not consider supersymmetry in this paper, and we focus on linking our modified hybrid inflation model to the EW vacuum stability in a non-SUSY case. However, if supersymmetry is considered as a UV completion to the SM, with a very large breaking scale, the EW vacuum stability is still questionable.} We will work in Planck units where the reduced Planck mass $M_{\text{P}}=1$.
The global minimum of the potential~(\ref{eq:vmhi}) is located at 
\begin{equation}
\langle\phi\rangle=0,\quad \langle\psi\rangle=\dfrac{v_\psi}{\sqrt{2}},\quad \langle\chi\rangle= \dfrac{v_\chi}{\sqrt{2}},
\end{equation} 
at which $V=0$. Since the inflaton $\phi$ acquires a zero vev, it does not mix with the other two fields $\psi,~\chi$ in the mass matrix and it is separated with a squared mass 
\begin{align}
m_\phi^2=m^2+2\lambda_{\phi \psi} v_{\psi}^2-2\lambda_{\phi \chi} v_{\chi}^2 .
\end{align}
On the other hand, the mass matrix in the basis $(\psi,\chi)$ is given by 
\begin{align}
{\cal M}_{\psi\chi}^2=4\begin{pmatrix}
\lambda_{\psi} v_{\psi}^2 & \lambda_{\psi \chi} v_{\psi} v_{\chi} \\
\lambda_{\psi \chi} v_{\psi}v_{\chi} & \lambda_{\chi} v_{\chi}^2 
\end{pmatrix}
\end{align}
with the following squared masses
\begin{equation}
m^2_{\psi,\chi}=2\Big[~\lambda_{\psi} v_{\psi}^2+\lambda_{\chi} v_{\chi}^2\pm\sqrt{(\lambda_{\psi} v_{\psi}^2-\lambda_{\chi} v_{\chi}^2)^2+4\lambda_{\psi \chi}^2 v_{\psi}^2 v_{\chi}^2}~\Big]
\end{equation}
It is clear that in the absence of the mixing $\lambda_{\psi \chi}$ between $\psi$ and $\chi$, the squared masses are given by $m^2_\psi=4\lambda_\psi v_\psi^2,~m^2_\chi=4\lambda_\chi v_\chi^2$.

The inflationary trajectory is obtained by minimizing the MHI potential~(\ref{eq:vmhi}) with respect to the $\psi$ and $\chi$ fields. It turns out that during inflation, $\psi$ is frozen at the origin while $\chi$ is shifted to a nonzero value on the trajectory 
\begin{equation}\label{eq:trajectory}
(\psi,\chi)=\left(0 \,,\, \sqrt{\frac{\lambda_{\phi \chi}}{\lambda_\chi} \phi ^2+\frac{\lambda_{\psi \chi}}{2\lambda_{\chi}} v_{\psi}^2+\frac{1}{2} v_{\chi}^2} ~ \right)
\end{equation}
Using the field-dependent squared mass matrix, a critical value $\phi_c$ that triggers the waterfall phase has the form
\begin{equation}
\phi_c= \frac{v_{\psi}}{\sqrt{2}} \sqrt{\frac{\lambda_{\psi} \lambda_{\chi} -\lambda_{\psi \chi}^2}{\lambda_{\chi} \lambda_{\phi \psi} + \lambda_{\phi \chi} \lambda_{\psi \chi}}} \hspace{0.5cm} \xrightarrow[]{(\lambda_{\psi\chi} \to 0)} \hspace{0.5cm} 
\frac{v_{\psi}}{\sqrt{2}}\sqrt{\frac{\lambda_\psi}{\lambda_{\phi\psi}}} 
\end{equation}
In that respect, the inflation effective potential takes the form
\begin{align}\label{eq:infpot}
V^{\rm inf}(\phi)=V_0+ \alpha \phi^2-\beta \phi^4,
\end{align}
with the following parameters
\begin{equation}\label{eq:shafipar1}
V_0= \dfrac{v_{\psi}^4}{4} \Big(\lambda_{\psi}-\dfrac{\lambda_{\psi \chi}^2}{\lambda_{\chi}}\Big), \quad
\alpha= \dfrac{m^2}{2}-\lambda_{\phi \chi} \Big(v_{\chi}^2+\dfrac{\lambda_{\psi \chi}}{\lambda_{\chi}}v_{\psi}^2\Big), \quad
\beta= \dfrac{\lambda_{\phi \chi}^2}{\lambda_{\chi}}.
\end{equation}
\begin{figure}[h!]
\begin{center}
\includegraphics[width=0.6\textwidth]{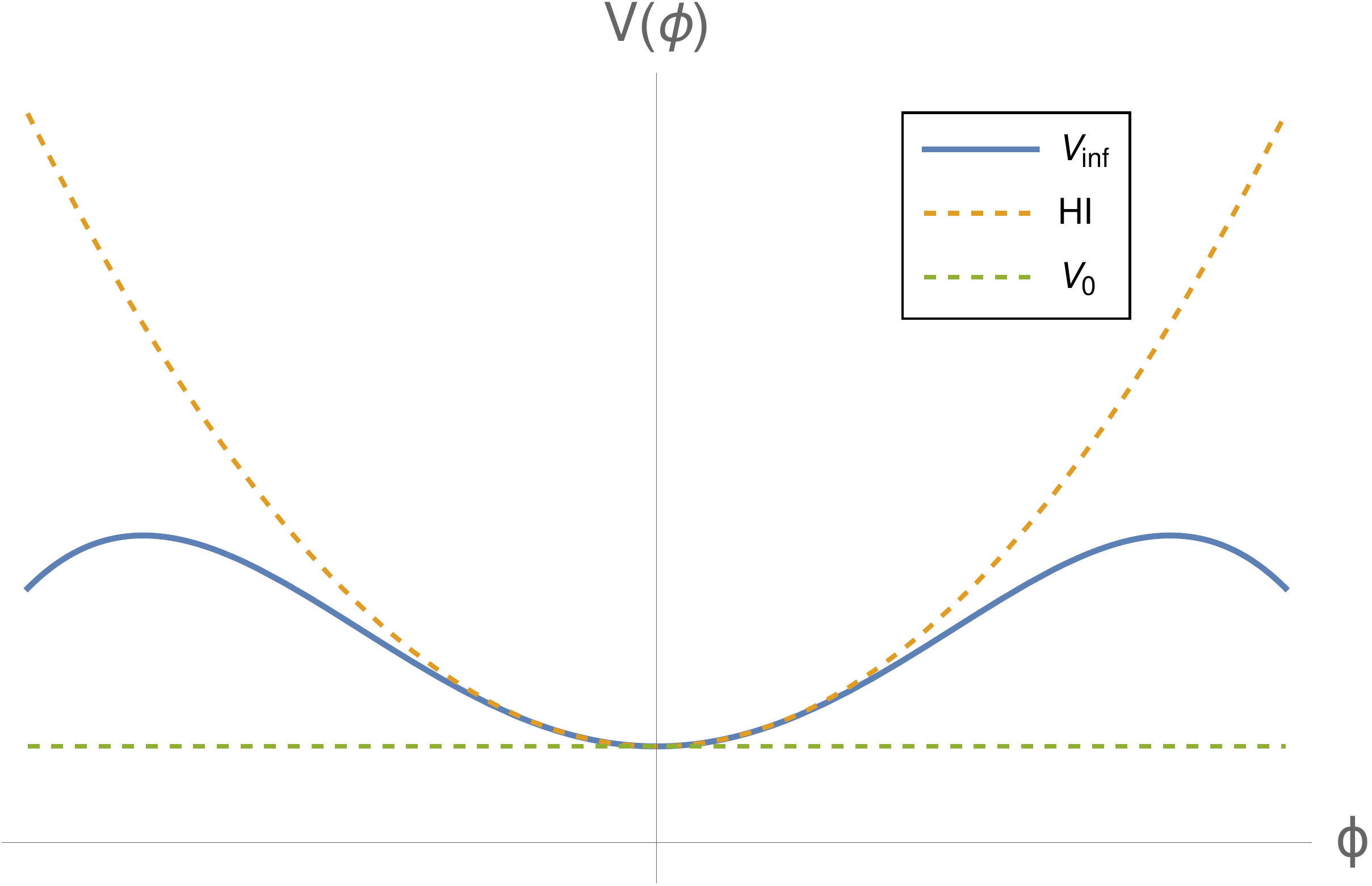}
\caption{\label{fig:infpots}The solid (blue) curve represents the MHI inflation potential~(\ref{eq:infpot}), while the dashed (yellow) curve represents the standard hybrid inflation potential.}
\end{center}
\end{figure} 
Here, the vacuum energy $V_0$ should be positive, hence $\lambda _{\psi}\lambda _{\chi}>\lambda _{\psi \chi}^2$. The latter condition is consistent with the requirement of having a real value for $\phi_c$. Since the coefficient $\beta$ is positive, with the negative sign term in~(\ref{eq:infpot}), there is a possibility that the inflaton rolls down from a hilltop close to $\phi=0$, if $\alpha$ is negative as well. This case will be similar to the inverted hybrid inflation~\cite{Lyth:1996kt}. However, the system, in this case, is unstable as $\phi$ should be larger than $\phi_c$, in order for the mass squared of $\psi$ to be positive during the inflation. Therefore, the only possibility is that $\alpha$ being positive, such that the potential shape is as illustrated in Fig.~\ref{fig:infpots} by the (blue) solid curve. Accordingly, the inflaton will roll down from large value close to the maximum, ${\phi}_{\rm m}$, towards the origin. This introduces a hilltop-type inflation~\cite{Boubekeur:2005zm,Kohri:2007gq}.\footnote{In Ref.~\cite{Rehman:2009wv}, it was indicated that one-loop quantum corrections to the hybrid inflation potential results in a hilltop-type potential. The dominant contribution comes from the RHN sector. This may be dangerous from the point of view of vacuum stability, which may become worse for large neutrino Yukawa couplings.}

It turns out that the potential curvature is negative around the maximum, hence the slow-roll parameter $\eta$ will be negative. This will alleviate the usual problem of large spectal index $n_s$, that is associated with hybrid inflation models. On the other hand, the tensor to scalar ratio $r$ will not be so small as the usual situation in the ordinary hybrid inflation.

It is customary to write the inflation potential in terms of the dimensionless parametrized field $\tilde{\phi}=\sqrt{\dfrac{\eta_0}{2}} \phi$, as follows
{
\begin{equation}\label{eq:mhiinfeffpot}
V^{\rm inf}(\tilde{\phi})=V_0\big( 1+ \tilde{\phi}^2- \gamma \tilde{\phi}^4\big), 
\end{equation}
}
where the dimensionful parameter $\eta_0$ and the dimensionless parameter $\gamma$ are
\begin{align}\label{eq:shafipar2}
\eta_0=\dfrac{2\alpha}{V_0},\quad \gamma=\frac{4\beta}{\eta_0^2V_0}.
\end{align}
Thus
\begin{align}
\tilde{\phi}_c=\sqrt{\dfrac{2\eta_0 V_0}{2\lambda_{\phi \psi}v_\psi^2+\eta_0\sqrt{\left(\lambda_\psi v_\psi^4-4V_0\right)\gamma\,V_0}}}.
\end{align}
In Sec.~\ref{sec:observables}, we use the following set of independent parameters: $\lambda_{\psi},\lambda_{\phi\psi},\lambda_{\phi\chi},v_{\psi},v_{\chi}$ and $V_0,\eta_0,\gamma$, after solving (\ref{eq:shafipar1}) and (\ref{eq:shafipar2}) for $\lambda_{\chi},\lambda_{\psi\chi}$ and $m$ in terms of $V_0,~\eta_0$ and $\gamma$.
\begin{figure}[h!]
\begin{center}
\includegraphics[width=0.47\textwidth]{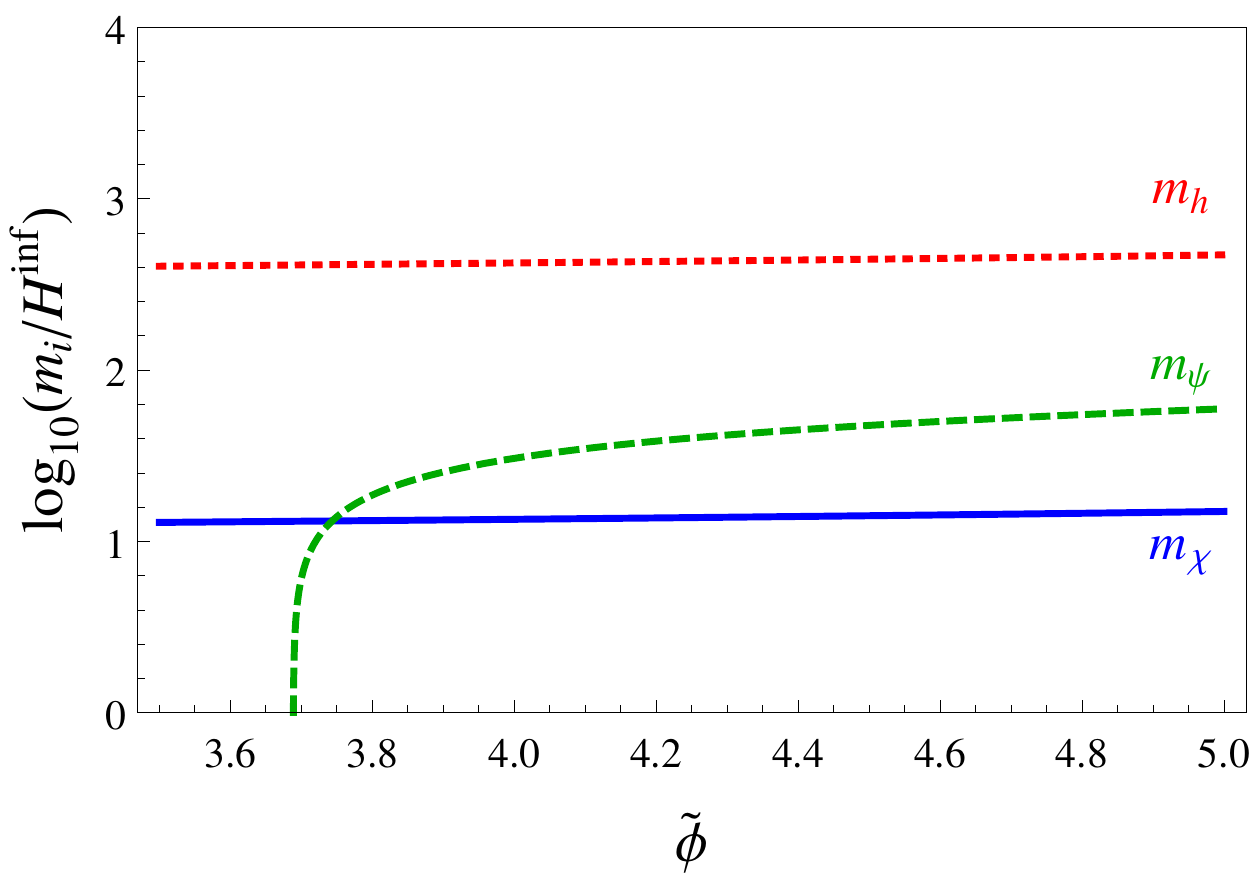}\quad
\includegraphics[width=0.47\textwidth]{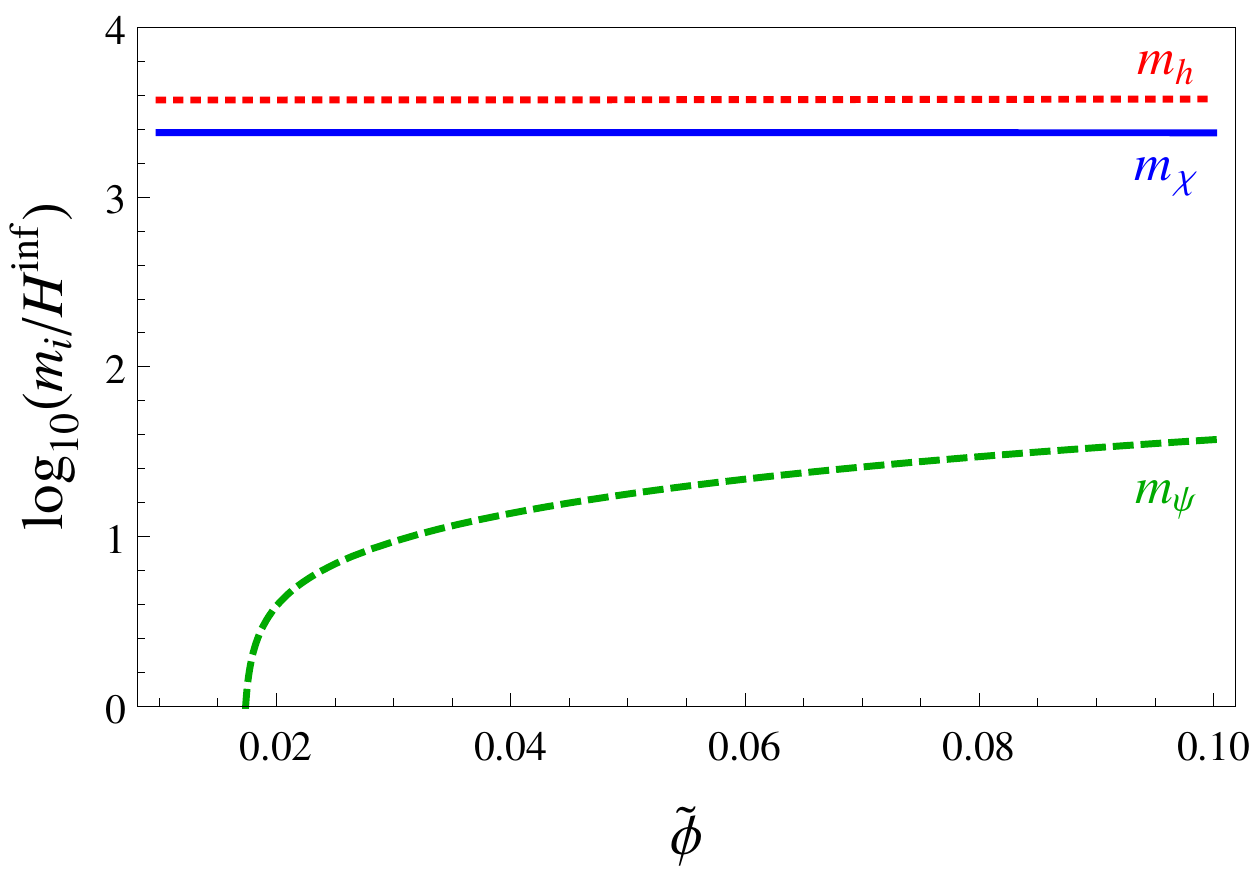}
 \caption{\label{fig:mhub} The logarithm of the ratios of the field-dependent masses of the heavy fields $m_{i}$: $m_{h}$ (red dotted), $m_\chi$ (blue solid), and $m_\psi$ (green dashed) to the Hubble scale $H^{\text{inf}}$ versus the inflaton field $\tilde{\phi}$ during inflation, for both the trans-Planckian BP1 (left panel) and the sub-Planckian BP2 (right panel) of Table~\ref{tab:BP1} and Table~\ref{tab:vsbp3}. }
\end{center}
\end{figure} 

The Hubble scale during inflation is given by
\begin{equation}\label{eq:mhiinfeffpot}
H^{\rm inf}(\tilde{\phi})\simeq \sqrt{\dfrac{V^{\rm inf}(\tilde{\phi})}{3}},
\end{equation}
 
Now, we move to discuss the stability of the inflation trajectory (\ref{eq:trajectory}). In Sec.~\ref{sec:vs}, we couple the SM Higgs, $h$, to the inflation sector as indicated in Eq.~(\ref{eq:V4}). Therefore, minimizing the total potential with respect to the SM Higgs field $h$ as well, implies that $h$ will be frozen at the origin during inflation with very large mass $m_h(\phi)$  compared to the Hubble scale as shown in Fig.~\ref{fig:mhub}. We demonstrate that only the direction of the inflaton field $\phi$ is light, whereas the other three fields $\psi$, $\chi$ and, $h$ are heavy during inflation. Figure~\ref{fig:mhub}, depicts the logarithm of the ratios of the field-dependent masses $m_i(\phi)$ of the heavy fields to the Hubble scale $H^{\text{inf}}(\phi)$ versus the inflaton field $\tilde{\phi}$, during inflation, for two benchmark points (BP1 and BP2  of Table~\ref{tab:BP1} and Table~\ref{tab:BPsinfobs}). Since the ratios are greater than 1, $\psi,~\chi$, and $h$ will not perturb the single field inflation dynamics in the $\phi$ direction, that is realized by the potential~\ref{eq:mhiinfeffpot}.

We conclude this section by discussing the end of inflation. When $\phi=\phi_c$, the waterfall phase starts, hence ending the inflation, and all fields stabilize at their true minimum values \cite{Linde:1993cn}. This assumption is valid if the waterfall phase happens in a short time $\Delta t \sim H^{-1}$, starting from the time $t_c$ at which $\phi=\phi_c$. In this case, $\phi$ rolls to the origin within a time duration much smaller than $H^{-1}$. It turns out that this is the case in our setup as well. We solve the equations of motion of the scalar fields 
\begin{equation}\label{eq:eqmotion}
\ddot{S_i} + 3 H \dot{S_i}+ \partial_i V_{\rm MHI}=0
\end{equation}
where $S_i$ denote all scalar fields in the inflation sector, $H=\dfrac{1}{\sqrt{3}M_{\text{P}}} \sqrt{\dfrac{1}{2}\sum_i  \dot{S_i}^2 + V_{\rm MHI}}  $ is the Hubble scale, $\partial_i$ is the derivative with respect to the scalars $S_i$ and we used the parameter values in Table~\ref{tab:BP1} and Table~\ref{tab:vsbp3}. We found that the time duration spent by the inflaton $\phi$ from $t_c$ to reach its minimum is  $ \ll H^{-1}$. Therefore, inflation ends once $\phi$ reaches $\phi_c$.

\section{\label{sec:observables}INFLATION OBSERVABLES}
In this section, we explore the parameter space that provides consistent inflation observables.
The slow roll parameters of inflation are given by
\begin{align}
\epsilon=\frac{\eta_0}{4}\left({\frac{V^{\rm inf}_{\tilde{\phi}}}{V^{\rm inf}}}\right)^2, 
\quad\quad\quad\quad
\eta=\frac{\eta_0}{2}\left({\frac{V^{\rm inf}_{\tilde{\phi}\tilde{\phi}}}{V^{\rm inf}}}\right),
\end{align}
and the number of $e$-foldings is given by
\begin{align}
N_e=\frac{1}{\sqrt{\eta_0}}\bigintsss_{\tilde{\phi}_{e}}^{\tilde{\phi}_*}{\frac{d\tilde{\phi}}{\sqrt{\epsilon(\tilde{\phi})}}}
\end{align}
where $\tilde{\phi}_*$ and $\tilde{\phi}_{e}=\tilde{\phi}_{c}$ are the field values at the time of horizon exit, and the end of inflation, respectively.
The spectral index $n_s$, the tensor to scalar ratio $r$ and the scalar amplitude $A_s$ are given respectively as 
\begin{align} 
n_s&=1- 6 \epsilon_* + 2 \eta_*=1-2\eta_0\frac{6\gamma^2{\tilde{\phi}_*}^6-5\gamma{\tilde{\phi}_*}^4+2(1+3\gamma){\tilde{\phi}_*}^2-1}{(1+{\tilde{\phi}_*}^2-\gamma{\tilde{\phi}_*}^4)^2} ,\\
r&=16 \epsilon_*=16 \eta_0\Big(\frac{{\tilde{\phi}_*}-2\gamma {\tilde{\phi}_*}^3}{1+{\tilde{\phi}_*}^2-\gamma{\tilde{\phi}_*}^4}\Big)^2 , \\
A_s&=\frac{V^{\rm inf}_*}{24 \pi^2 \epsilon_*}=\frac{V_0(1+{\tilde{\phi}_*}^2-\gamma{\tilde{\phi}_*}^4)^3}{24 \pi^2 \eta_0({\tilde{\phi}_*}-2\gamma {\tilde{\phi}_*}^3)^2} , 
\end{align}
where the subscript $``*"$ means that all quantities are calculated at horizon exit time.
\begin{figure}[h!]
\begin{center}
\includegraphics[width=0.48\textwidth]{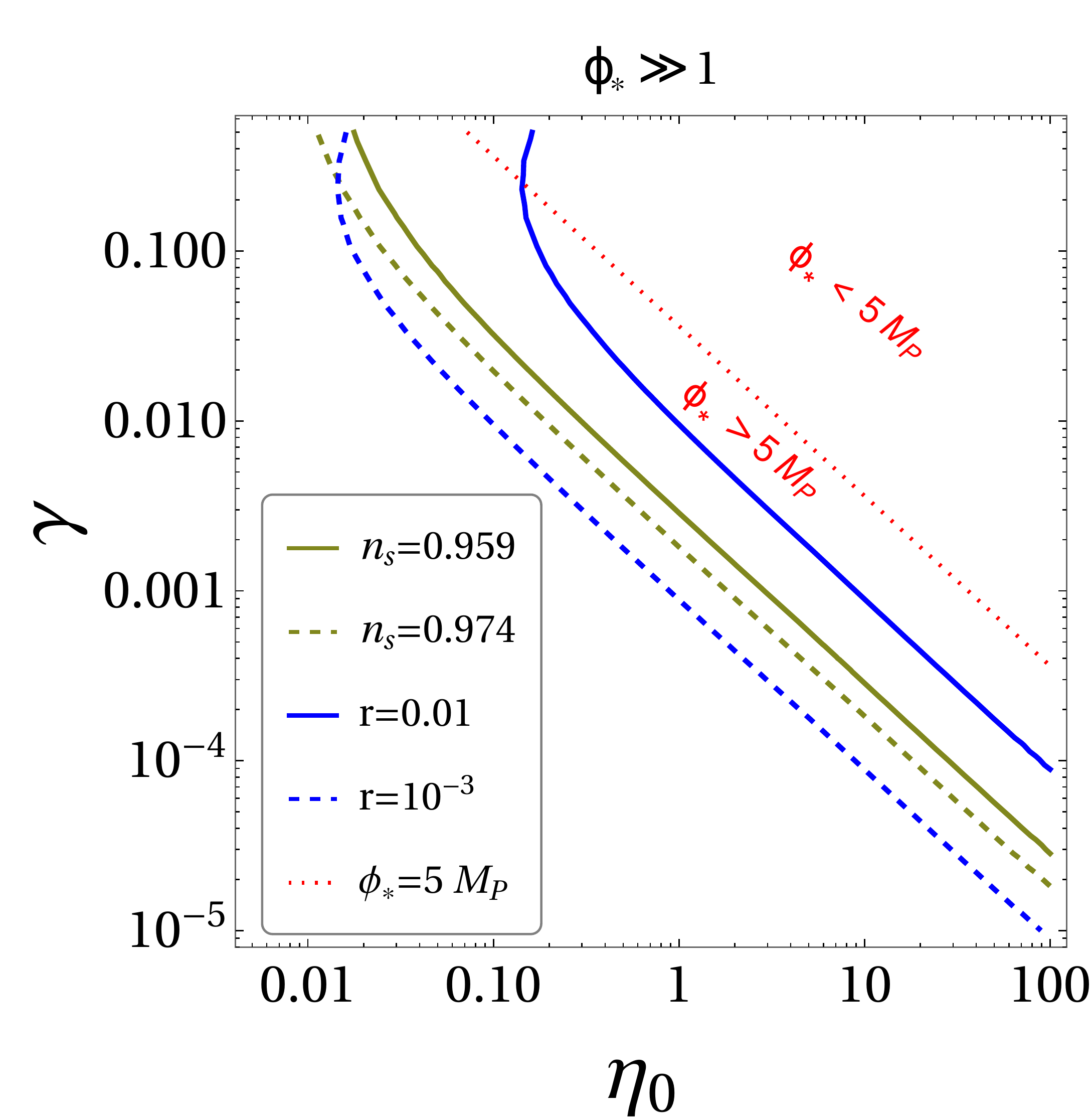}
\includegraphics[width=0.45\textwidth]{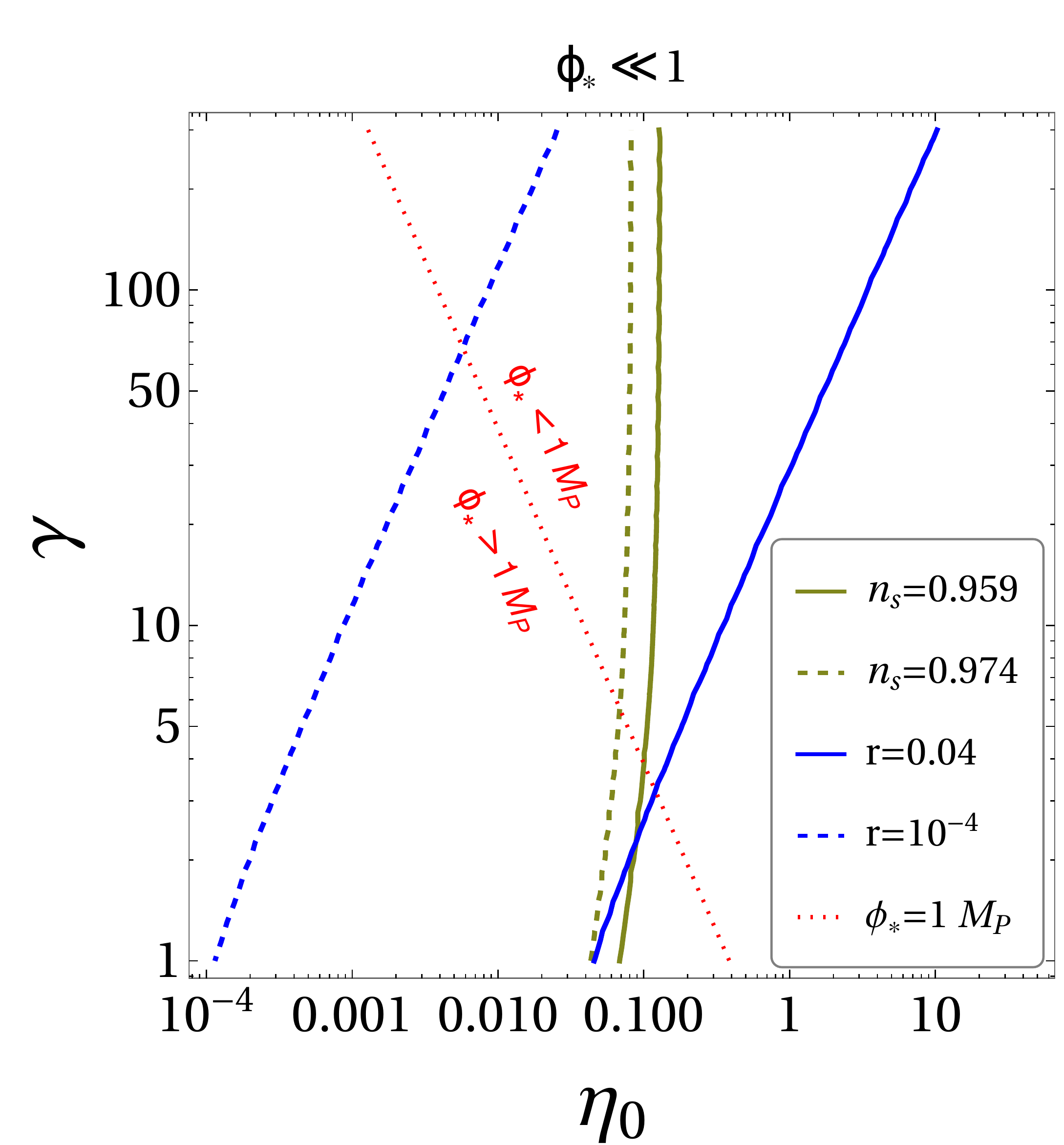}
\caption{\label{fig:eta0-gamma}Contours in the $\eta_0$--$\gamma$ plane corresponding to different values of $n_s$ and $r$, for both cases of large field inflation (left panel) and small field inflation (right panel).}
\end{center}
\end{figure} 

As advocated in the previous section, in order to have successful inflation, the condition $\tilde{\phi}_c < \tilde{\phi}_*< \tilde{\phi}_{\rm m}=\dfrac{1}{\sqrt{2\gamma}}$, should be fulfilled. Hence we differentiate between two cases: when $\tilde{\phi}_* \gg 1 $, (corresponding to $\gamma\ll 1)$, and $\tilde{\phi}_* \ll 1$, (corresponding to $\gamma\gg 1)$, where $\tilde{\phi}_{*}$ value is close to $\tilde{\phi}_{\rm m}$ value. It turns out that the field excursion during inflation occurs mostly in the negative curvature part of the potential (with negative $\eta$), before the inflaton reaches the inflection point of the potential, which reduces $n_s$.

\paragraph{\textit{Case I: Large $\tilde{\phi}_*$ regime (trans-Planckian).}}
We expand for large $\tilde{\phi}_*$, after expressing $\gamma$ in terms of $\tilde{\phi}_m$ that is close to $\tilde{\phi}_*$. Therefore, the slow-roll parameters can be approximated as
\begin{align} 
\epsilon \simeq \dfrac{16 \eta_0}{\tilde{\phi}_*^4} \big( \tilde{\phi}_{\rm m}- \tilde{\phi}_* \big)^2 ,
\quad\quad\quad\quad
\eta\simeq - \dfrac{4 \eta_0}{\tilde{\phi}_*^2} + 
\dfrac{20 \eta_0}{\tilde{\phi}_*^3} \big( \tilde{\phi}_{\rm m}- \tilde{\phi}_* \big),
\end{align}
hence,  we have
\begin{align} 
n_s \simeq 1- \dfrac{8 \eta_0}{\tilde{\phi}_*^2}+ 
\dfrac{40 \eta_0}{\tilde{\phi}_*^3} \big( \tilde{\phi}_{\rm m}- \tilde{\phi}_* \big), 
\quad\quad\quad\quad
r\simeq \dfrac{256 \eta_0}{\tilde{\phi}_*^4} \big( \tilde{\phi}_{\rm m}- \tilde{\phi}_* \big)^2 .
\end{align}
In this case, it is estimated that $0.05\lesssim\eta_0 \lesssim 31$, for changing $\gamma \in [10^{-4},0.25]$. In this regard, a correlation between $\gamma$ and $\eta_0$ is shown in the left panel of Fig.~\ref{fig:eta0-gamma}, where the intersection region between the four curves is allowed by the observational constraint on $n_s$ and $r$. In the latter figure, the contours correspond to different values of $n_s$ and $r$, where $\tilde{\phi}_*$ values were chosen to be close to $ \tilde{\phi}_{\rm m}= \dfrac{1}{\sqrt{2\gamma}}$. This case represents large field inflation, where $\phi_*$ values are trans-Planckian as indicated by the region below the red dotted curve that corresponds to $\phi_*= 5 \, M_{\rm P}$. 

Figure~\ref{fig:eta0-gamma} (left panel) implies that $\gamma$ decreases as $\eta_0$ increases. For very small values of $\gamma$, $\eta_0$ becomes very large hence $\tilde{\phi}_*$ is very large. We may control the value of $v_\chi$ in order to reduce $m_\chi \lsim \Lambda_\chi $ and make use of the threshold effect and stabilize the EW vacuum. We may consider the $\chi$ being the SM-like Higgs if $v_\chi \sim 10^{-16}$ as in BP3 of Table~\ref{tab:BP1}. Accordingly, the SM-like Higgs with squared mass eigenvalue is given as
\begin{equation}
m_\chi^2 \simeq 4v_{\chi}^2 \Big(\lambda_{\chi}-\dfrac{\lambda_{\psi \chi}^2}{\lambda_{\psi}}\Big).
\end{equation}
However, we need to reduce $m_\psi \sim \Lambda_I$, with threshold correction given by $\dfrac{\lambda_{\psi \chi}^2}{\lambda_{\psi}} \gtrsim 0.1$~\cite{Elias-Miro:2012eoi}, hence $\lambda_\chi > 0.1$ guarantees real $\phi_c$ and $m_\chi$.
In this case, $V_0$ is very small and insufficient for accounting for the inflation energy scale and $A_s$ limits. 
However, trans-Planckian values of $\phi_*$ can account for the inflation obesrvables and the inflation scale is fixed by $m$ as indicated by the benchmark point BP3 in Table~\ref{tab:BP1} and Table~\ref{tab:BPsinfobs}. In fact, the only problem of this scenario is that the inflation will not end after $60$~$e$-folds since $\phi_c$ is always small, constrained by the value of $m_\psi \sim \Lambda_I $.
\begin{table}[h]
\begin{subtable}{1.\linewidth}
\begin{center}
{\begin{tabular}{cccccc} 
\hline
\hline
{Par} & $\lambda_{\psi}$ & $\lambda_{\chi}$ & $\lambda_{\phi\psi}$ & $\lambda_{\phi\chi}$ & $\lambda_{\psi\chi}$ \\\hline
{BP1} & $1.00\times 10^{-3}$ & $7.15\times 10^{-8}$ & $9.80\times 10^{-10}$ & $1.5\times 10^{-11}$ & $3.48\times 10^{-7}$ \\\hline
{BP2} & $3.75\times 10^{-7}$ & $3.30\times 10^{-2}$ & $3.75 \times 10^{-8}$ & $1.60\times 10^{-7}$ & $1.11\times 10^{-4}$ \\\hline
{BP3} & $1.60 \times 10^{-6}$ & $5.89 \times 10^{-3}$ & $3.41 \times 10^{-21}$ & $1.80\times 10^{-9}$ & $2.66 \times 10^{-5}$ \\\hline
\hline
\end{tabular}}
\caption{\label{tab:BP1-1}Benchmark points (BPs) for dimensionless couplings of the MHI potential~(\ref{eq:vmhi}) which produce the observables in Table~\ref{tab:BPsinfobs}.}
\end{center}
\end{subtable}
\\\vspace{0.3cm}\\
\begin{subtable}{1.\linewidth}
\begin{center}
{\begin{tabular}{cccccc} 
\hline
\hline
{Par} & $m$ & $v_{\psi}$ & $v_{\chi}$ & ${\phi}_{*}$ & ${\phi}_{c}$ \\\hline
{BP1} & $2.24\times 10^{-6}$ & $1.862\times 10^{-2}$ & \begin{tabular}{c}$1.30\times 10^{-7}$\\\ $4.16 \times 10^{-8}$\end{tabular} & $17.0596$ & $12.8191$ \\\hline
{BP2} & $2.48\times 10^{-5}$ & $7.52\times 10^{-1}$ & \begin{tabular}{c}$5.00\times 10^{-9}$
\\
$1.70\times 10^{-10}$\end{tabular} & $0.97000$ & $0.11403$ \\\hline
{BP3} & $9.67\times 10^{-7}$ & $3.16 \times 10^{-6}$ & $3.53 \times 10^{-16}$ & $19.4091$ & $6.75\times 10^{-5}$ \\\hline
\hline
\end{tabular}}
\caption{\label{tab:BP1-2}BPs for dimensionful parameters (in $M_{\rm P}$) of the MHI potential~(\ref{eq:vmhi}) which in addition to (Table~\ref{tab:BP1-1}), produce the observables in Table~\ref{tab:BPsinfobs}.}
\end{center}
\end{subtable}
\\\vspace{0.3cm}\\
\begin{subtable}{1.\linewidth}
\begin{center}
{\begin{tabular}{cccccc} 
\hline
\hline
{Par} & $V_0~(M^4_{\text{P}})$ & $\eta_0~(M_{\text{P}}^{-2})$ & $\gamma$ & $\tilde{\phi}_{*}$ & $\tilde{\phi}_{c}$ \\ \hline
{BP1} & $3.00\times 10^{-11}$ & $1.65\times 10^{-1}$ & $1.54\times 10^{-2}$ & 4.90 & 3.682 \\\hline
{BP2} & $1.40\times 10^{-10}$ & $4.67\times 10^{-2}$ & $10.15$ &$1.48\times 10^{-1}$ &$1.74\times 10^{-2}$ \\\hline
{BP3} & $3.70\times 10^{-29}$ & $2.5\times 10^{16}$ & $9.5\times 10^{-20}$ &$2.17 \times10^{9}$ &$7.547 \times10^{4}$ \\\hline
\hline
\end{tabular}}
\caption{\label{tab:BP1-3}BPs of the effective inflation potential~(\ref{eq:mhiinfeffpot}) corresponding to parameters' values in Table~\ref{tab:BP1-1} and Table~\ref{tab:BP1-2} which produce the observables in Table~\ref{tab:BPsinfobs}.}
\end{center}
\end{subtable}
\caption{\label{tab:BP1}BPs of the MHI model which produce the observables in Table~\ref{tab:BPsinfobs}.}
\end{table}
\begin{table}[h]
\begin{center}
{\begin{tabular}{cccccccc} 
\hline
\hline
{Obs} & $N_e$ & $n_s$ & $r$ & $A_s$ & $m_\phi$ & $m_\psi$ & $m_\chi$ \\\hline
{BP1} & 59.6 & 0.9688 & 0.0165 & $1.98\times 10^{-9}$ & $5.80\times 10^{12}$ & $2.86\times 10^{15}$ & \begin{tabular}{c}$1.41\times 10^{8}$\\ $5.25 \times 10^{7}$\end{tabular} \\\hline
{BP2} & 52.3& 0.9669 & 0.0049 & $1.97\times 10^{-9}$ & $4.97 \times 10^{14}$ & $2.21 \times 10^{15}$ & \begin{tabular}{c}$2.97\times10^8$\\
$1.07 \times 10^{7}$\end{tabular}
\\
\hline
{BP3} & 162.2 & 0.9664 & 0.0030 & $2.10\times 10^{-9}$ & $2.34\times 10^{12}$ & $1.94\times 10^{10}$ & 125.06 \\\hline
\hline
\end{tabular}}
\caption{\label{tab:BPsinfobs}Inflation observables and scalar masses (in GeV) corresponding to the parameters values given in Table~\ref{tab:BP1}.}
\end{center}
\end{table}
\begin{figure}[h]
\begin{center}
\includegraphics[scale=0.6]{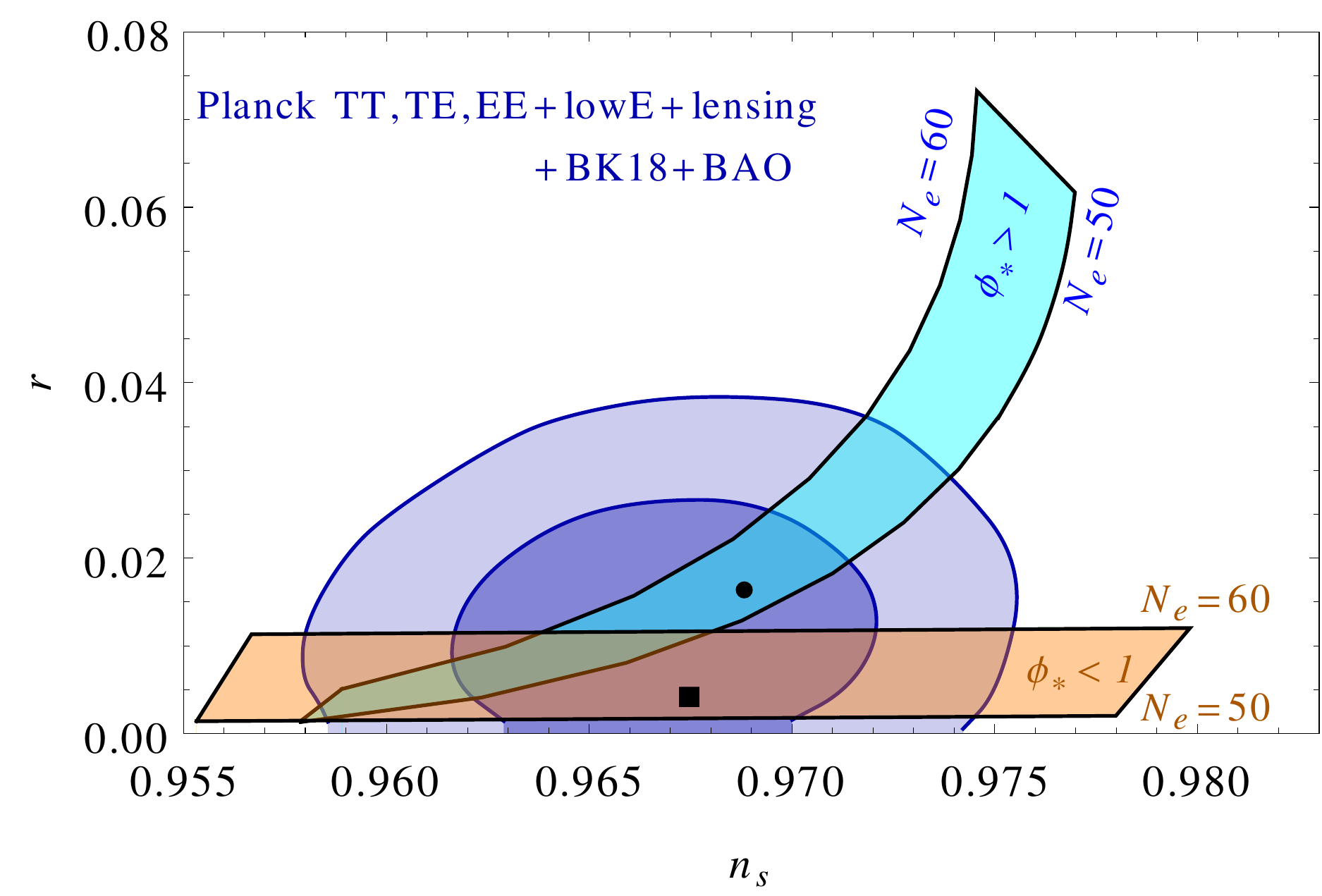}
\caption{\label{fig:planck}Predictions of the MHI model in the $(n_s,r)$ plane are given by the cyan patch (trans-Planckian) and the orange patch (sub-Planckian). The blue contours are the observed constraints extracted from Planck 2018, and they correspond to the observed $68\%$ and $95\%$ C.L. constraints in $(n_s,r)$ plane when adding BICEP/Keck and BAO data~\cite{Planck:2018jri,BICEP:2021xfz}. The two BPs indicated by the solid dot and square are BP1 and BP2 presented in Table~\ref{tab:BP1}.}
\end{center}
\end{figure} 

\paragraph{\textit{Case II: Small $\tilde{\phi}_*$ regime (sub-Planckian).}}
If $\gamma \gg 0.5 $, then $\tilde{\phi}_* \ll 1$. 
Expanding for small $\tilde{\phi}_*$, the slow-roll parameters $\epsilon$ and $\eta$ can be approximated as 
\begin{align}
\epsilon\simeq \eta_0 \tilde{\phi}_*^2,
\quad\quad\quad\quad
\eta \simeq \eta_0 - \left(6 \gamma +1 \right) \eta_0\tilde{\phi}_*^2 \,,
\end{align}
hence $n_s$ and $r$ have the following approximated forms:
\begin{align}
n_s\simeq 1 + 2 \eta_0 -4 \left(3 \gamma +2 \right) \eta_0\tilde{\phi}_*^2,
\quad\quad\quad\quad
r \simeq 16 \eta_0 \tilde{\phi}_*^2 \, .
\end{align}
This case corresponds to sub-Planckian values of $\phi_*$ as shown in Fig.~\ref{fig:eta0-gamma}, right panel, for the region above the red dotted curve. In Table~\ref{tab:BP1} and Table~\ref{tab:BPsinfobs}, we give one benchmark point (BP2) that accounts for this case. It turns out that $v_\chi $ is not constrained, so we have chosen it such that $m_\chi \sim \Lambda_I$ hence it can modify the SM RGEs rendering the EW vacuum stable. 

In Fig.~\ref{fig:planck}, we present the predictions of our model for both cases of large and small field regimes, versus the observed constraints on $n_s$ and $r$ by~\cite{BICEP:2021xfz}. The cyan patch presents the large field inflation predictions. We scanned over $\gamma\in[0.001,0.1]$, $\eta_0\in[0.18,0.21]$ and $\tilde{\phi}_{*} \in[5,6]$. The black curves correspond to fixed $\eta_0$, $\tilde{\phi}_{*}$ and number of $e$-folds for each, namely, $N_e=50,\eta_0=0.18$ $\tilde{\phi}_{*}$=5.25 and $N_e=60,\eta_0=0.21$, $\tilde{\phi}_{*}$=5.4, while $\gamma$ changes along each curve. On the other hand, the orange patch represents the small-field inflation predictions. We scanned here over $\gamma\in[5,100]$, $\eta_0\in[0.01,0.1]$ and $\tilde{\phi}_{*} \in[0.090,0.099]$. The black lines correspond to fixed $\eta_0$, $\tilde{\phi}_{*}$ and number of $e$-folds for each, with $N_e=50,\eta_0=0.04$ $\tilde{\phi}_{*}$=0.098 and $N_e=60,\eta_0=0.2$, $\tilde{\phi}_{*}$=0.0945, while $\gamma$ changes along each line.
The observed value of the scalar amplitude $A_s \simeq 2.1\times 10^{-9}$, fixes $V_0\sim 10^{-11}- 10^{-10}$ in both cases of large and small field inflation. The solid (black) dot and square correspond to parameters' values and obervables given in Table~\ref{tab:BP1} and~\ref{tab:BPsinfobs} for BP1 and BP2, respectively. 
\section{\label{sec:reheat}REHEATING AND QUANTUM CORRECTIONS}
In this section, we discuss the reheating where the inflation decays into RH neutrinos $N$ and possible quantum corrections that may modify the inflation effective potential. The complete Lagrangian that is responsible for neutrino masses and reheating contains the SM Higgs $h$ and left-handed neutrinos $\nu_L$ and has the form
\begin{align}\label{eq:rhnlag}
{\cal L}_{\nu}= Y_\nu \, h \, \bar{\nu}_L \, N + Y_\phi \, \phi \, \bar{N	} \, N + Y_\psi \, \psi \, \bar{N} \, N + Y_\chi \, \chi \, \bar{N} \, N + m_N \, \bar{N} \, N 
\end{align}
where $m_N$ is a mass scale that is constrained by reheating and the seesaw mechanism of generating the neutrino masses. Accordingly, the tiny neutrino masses are given by
\begin{equation}\label{eq:mnu}
m_\nu= \dfrac{Y_\nu^2 \, v^2}{M_N},
\end{equation}
where $M_N= m_N + Y_\psi \, v_\psi + Y_\chi v_\chi $. The reheating temperature is given by~\cite{Ellis:2015jpg,Ellis:2016spb,Dudas:2017rpa,Khlopov:1993ye,Khlopov_2006,osti_6298609,KHLOPOV1984265}
\begin{equation}
T_{\rm R} =\left( \dfrac{40}{g_s \pi^2} \right)^{1/4} \, \sqrt{\Gamma \, M_{\rm P}}
\end{equation}
where $g_s$ is the effective number of light degrees of freedom in a thermal bath at temperature $T_{\rm R}$ and $\Gamma$ is the total decay width of the inflation sector fields that is given by
\begin{align}
\Gamma= \sum_{i=1}^3 \Gamma_{i \to N N}= \sum_{i=1}^3 \dfrac{Y_i^2 \, m_i}{8\pi} ,
\end{align}
where $i$ runs over $S_i=\phi,\psi,\chi $ and the decay $\Gamma_{i \to N N}$ is kinematically allowed if $m_i > 2 M_N$. Our model is non-SUSY, so gravitino overproduction constraints on $T_{\rm R}$, are not applicable here.\footnote{See Ref.~\cite{Moursy:2021kst} and other references therein.} We will consider $T_{\rm R} \lsim 2\times 10^{11}$ GeV.
In case of large field inflation (BP1 in Table~\ref{tab:BP1}), kinematically allowed decays for reheating implies $Y_\phi \lsim 4.77\times 10^{-4}$, $Y_\psi \lsim 1.9\times 10^{-5}$ and $Y_\chi \lsim 1$. Accordingly, $M_N\simeq m_N + 2.7 \times 10^{11} $ GeV, hence $m_\nu \sim 0.1 $ eV implies that $Y_\nu \sim 0.03$, if $m_N$ is subdominant. There is no worry regarding reducing the SM EW instability scale $\Lambda_I$, as the contribution to the beta function of the SM Higgs quartic coupling is proportional to $Y_\nu^4\sim 10^{-6}$, which is negligible compared to the top quark Yukawa coupling contribution. Decay kinematics constrains $m_N \lsim 10^{15}$ GeV. If $m_N$ is dominant, then $Y_\nu$ can be $ {\cal O}(1)$. In this case, $m_\chi$ should be reduced to $ m_\chi \lesssim \Lambda_I\sim 10^5$ GeV. In that case, one solution is to reduce $m_N$ such that $Y_\nu \ll 1$ with negligible effects on the RGEs. 

On the other hand, for small field inflation (BP2 in Table~\ref{tab:BP1}), a reheating temperature $T_{\rm R} \lsim 1\times 10^{13}$ GeV, implies $Y_\phi \lsim 1.5\times 10^{-4}$, $Y_\psi \lsim 5 \times 10^{-3}$ and $Y_\chi \lsim 1$. Accordingly, $M_N\simeq m_N + 3.8 \times 10^{14} $ GeV, and seesaw mechanism fixes $Y_\nu \sim {\cal O}(1)$. This will affect the instability scale $\Lambda_I\sim 10^5$ GeV. Therefore, we need to reduce $ m_\chi \lesssim \Lambda_I$. In both cases, if $T_{\rm R}$ is reduced sufficiently, then the value of $Y_\nu $ will be reduced with negligible contributions to the RGEs.
However, the case of sizable values of $Y_\nu$ may be interesting, from the point of view of low-energy phenomenology signatures.

\paragraph{Quantum corrections}
The Lagrangian (\ref{eq:rhnlag}) implies an additional contribution to the inflation potential via the one-loop corrections~\cite{Coleman:1973jx}. Following Ref.~\cite{Rehman:2009wv}, the inflation potential~(\ref{eq:mhiinfeffpot}) will have the following form, after including the quantum correction
\begin{equation}
{\cal V}^{\rm inf}(\tilde{\phi})=V_0\left( 1+ \tilde{\phi}^2- \tilde{\gamma} \, \tilde{\phi}^4\right),
\end{equation}
where $ \tilde{\gamma}= \gamma + \dfrac{4}{V_0 \, \eta_0^2} \dfrac{Y_\phi^4}{16 \pi^2}\log\Big( \dfrac{\tilde{\phi}}{\tilde{\phi}_c}\Big)$. An upper bound on the reheating temperature $T_{\rm R} \lsim 2\times 10^{11}~(1\times 10^{13})$ GeV, for large~(small) field inflation, results in tiny contributions from quantum corrections to the inflation potential and the inflation observables will not alter. Beyond these constraints, quantum corrections have substantial contributions. In this case, inflation observables discussed in the previous section will not hold. However, adjusting $\eta_0$ can take us back to the observables given in the previous section.

\section{\label{sec:vs}HIGGS VACUUM STABILITY}
The SM Higgs doublet couples to the singlet scalar fields of the MHI potential~(\ref{eq:vmhi}) in the full scalar potential\footnote{\label{ftnt:lmphchsign}In Sec.~\ref{sec:vs}, each $\lambda_{\phi\chi}$ must be replaced with $-\lambda_{\phi\chi}$ for the negative sign convention of the MHI potential~(\ref{eq:vmhi}).}
\begin{align}
V(H,\phi,\psi,\chi)&= V_{\rm MHI}(\phi,\psi,\chi)+\lambda_h\Big(h^2-\frac{v^2}{2}\Big)^2\n 
&+2 \Big(h^2-\frac{v^2}{2}\Big) \Big[\lambda_{h\phi}\phi^2+\lambda_{h\psi}\Big(\psi^2-\frac{v_\psi^2}{2}\Big)+\lambda_{h\chi} \Big(\chi^2-\frac{v_\chi^2}{2}\Big)\Big].
\label{eq:V4}
\end{align}

\subsection{Matching conditions}
Following the procedure of Ref.~\cite{Elias-Miro:2012eoi}, at high scale $\sim\mathcal{O}(\text{GUT})$, the superheavy field $\psi$ is integrated out via its equation of motion and considering that it is a background static and homogeneous field. By neglecting its kinetic term ($\frac{\partial V}{\partial \psi}=0$), the field $\psi$ is given in terms of the other fields $h,~\chi$ and $\phi$ as
\begin{align}\label{eq:psi2phichih}
\psi^2&= \frac{v_{\psi}^2}{2}-\frac{\lambda_{\phi\psi}}{\lambda_{\psi}}\phi^2-\frac{\lambda_{\psi\chi}}{\lambda_{\psi}}\chi^2-\frac{\lambda_{h\psi}}{\lambda_{\psi}} h^2.
\end{align}
At a lower energy scale, the potential can be read off by substituting Eq.~(\ref{eq:psi2phichih}) into the full potential~(\ref{eq:V4}). We find that the effective 3-field potential of $h,~\phi$, and $\chi$ contains the following quartic terms which are relevant to the EW vacuum stability:
\begin{align}
 V_{3{\rm eff}}(h,\phi,\chi) & \supset\lambda_{3h} h^4 +\lambda_{3\phi} \phi^4 +\lambda_{3\chi}\chi^4+2h^2 [\lambda_{3h\phi}\phi^2+\lambda_{3h\chi} \chi^2]+2 \lambda_{3\phi\chi}\phi^2\chi^2,
 \label{eq:V3}
\end{align}
where the numerical values of the running couplings are matched to their corresponding ones in the full potential at the heavy mass $m_\psi$ scale threshold such that at the $\psi$ integration out boundary via the following matching conditions (MCs)
\begin{align}
 \label{eq:V3mtchngi1}
 \lambda_{3S}&= \lambda_{S}-\frac{\lambda^2_{S\psi}}{\lambda_{\psi}},\quad S=h,\chi,\phi\\
 \lambda_{3S_1S_2}&= \lambda_{S_1S_2}-\frac{\lambda_{S_1\psi}\lambda_{S_2\psi}}{\lambda_{\psi}},\quad S_1=h,\phi,\quad S_2=\phi,\chi,\quad S_1\neq S_2.
 \label{eq:V3mtchngi2}
\end{align}
where each $\lambda_{\phi\chi}$ must be replaced with $-\lambda_{\phi\chi}$ as mentioned in footnote~\ref{ftnt:lmphchsign}.\footnote{It is worth noticing that due to the $\phi$-$\psi$ interaction in~(\ref{eq:vmhi}), when $\psi$ is integrated out at $m_\psi$, an effective self-interaction quartic term $\lambda_{3\phi}\, \phi^4$ is generated at a low scale below $m_\psi$ due to the form of the $\psi$-integration out solution of Eq.~(\ref{eq:psi2phichih}). This $\phi^4$ term was avoided in the inflation potential~(\ref{eq:V4}) in order not to spoil the inflation as mentioned above after~(\ref{eq:vmhi}). Since $\lambda_{4\phi}\equiv\lambda_\phi\equiv0$, we see from Eq.~(\ref{eq:V3mtchngi1}) that the MC of the $\phi$ field quartic coupling reads $\lambda_{3\phi}=-\frac{\lambda^2_{\phi\psi}}{\lambda_{\psi}}$. Despite the fact that this condition cannot be verified for real values of parameters, with $\lambda_{3\phi},\lambda_\psi>0$ as demanded by the boundedness from the below conditions of the potential at different scales, we should not worry about it, since the running of $\lambda_{3\phi}$ is only up to $m_\psi$, while $\lambda_{\phi\psi}$ and $\lambda_{\psi}$ run starting only from $m_\psi$, and the boundary condition $\lambda_{3\phi}=-\frac{\lambda^2_{\phi\psi}}{\lambda_{\psi}}$ at $m_\psi$ is either verified for tiny values of $\lambda_{3\phi}\approx\frac{\lambda^2_{\phi\psi}}{\lambda_{\psi}}\approx0$ or $\lambda_{3\phi}$ is discontinuous at the $\Lambda_{3I}$ threshold for $\lambda_{3\phi}\gsim0$ and $\lambda_{\phi}(t)\equiv0$ for $t=\log_{10}{(Q/{\rm GeV})}\geq\Lambda_{3I}$ and this does not affect the running of the Higgs coupling at all scales and the EW vacuum stability, as clarified below.}
After that, at the lower scale $m_\phi\ll m_\psi$, the second heaviest field $\phi$ is integrated out by considering that $\frac{\partial V_{3{\rm eff}}}{\partial \phi}=0$, where we find 
\begin{equation}
 \phi^2=-\frac{m_3^2}{4\lambda_{3\phi}}-\frac{\lambda_{3h\phi}}{\lambda_{3\phi}} h^2-\frac{\lambda_{\phi\chi}}{\lambda_{3\phi}}\chi^2,
\end{equation}
and the effective 2-field potential of $h,\chi$ includes the following quartic terms:
\begin{equation}
 V_{2{\rm eff}}(h,\chi)\supset\lambda_{2h}h^4 +\lambda_{2\chi} \chi^4 +2\lambda_{2h\chi} h^2\chi^2.
 \label{eq:V2}
\end{equation}
The matching conditions of the couplings at the $m_\phi$ threshold are 
\begin{align}
 \label{eq:V2mtchngi1}
 \lambda_{2S}&= \lambda_{3S}-\frac{\lambda^2_{3S\phi}}{\lambda_{3\phi}},\quad S=h,\chi\\
 \lambda_{2h\chi}&= \lambda_{3h\chi}-\frac{\lambda_{3h\phi}\lambda_{3\phi\chi}}{\lambda_{3\phi}}.
 \label{eq:V2mtchngi2}
\end{align}

Finally, the remaining $\chi$ field is integrated out at the SM instability scale $\sim m_\chi\sim\mathcal{O}(10^8)$ GeV and the SM Higgs quartic coupling is modified at this scale as in the following matching condition
\begin{align}
\label{eq:V1mtchngi}
\lambda\equiv\lambda_{\text{SM}}&= \lambda_{2h}-\frac{\lambda^2_{2h\chi}}{\lambda_{2\chi}}.
\end{align}
Generally, when a field $\bar{S}=\phi,\psi,\chi=(S_1,S_2,S_3)$ is integrated out, the parameters of the remaining fields $S,S_1,S_2=(h,\phi,\psi,\chi)$ have the following MCs at the $m_{\bar{S}}$ scale
\begin{align}
\label{eq:genmtchngi1}
\lambda_{iS}&= \lambda_{i+1,S}-\frac{\lambda^2_{i+1,S\bar{S}}}{\lambda_{i+1,\bar{S}}},\quad S\neq\bar{S}\\
\lambda_{iS_1S_2}&= \lambda_{i+1,S_1S_2}-\frac{\lambda_{i+1,S_1\bar{S}}\lambda_{i+1,S_2\bar{S}}}{\lambda_{i+1,\bar{S}}},
\label{eq:genmtchngi2}
\end{align}
where each $\lambda_{\phi\chi}$ must be replaced with $-\lambda_{\phi\chi}$ as mentioned in footnote~\ref{ftnt:lmphchsign}.

\subsection{Renormalization group equations}
The relevant one-loop renormalization group equations (RGEs) of the Higgs and $S_i$'s quartic coupling take the form (for $S,S_1,S_2=\phi,\psi,\chi,~i=2,3,4$)
\begin{align}
\label{eq:rgeh}
\frac{16\pi^2}{\log(10)}\frac{d\lambda_{ih}}{dt}&= \beta_{ih}=\beta_{h}^{\text{SM}}+\beta_{h}^{\text{int}},\\
\label{eq:rgehf}
\frac{16\pi^2}{\log(10)}\frac{d\lambda_{ihS}}{dt}&= \beta_{ihS}=\beta_{hS}^{\text{SM}}+\beta_{hS}^{\text{int}}, \\
\label{eq:rgef}
\frac{16\pi^2}{\log(10)}\frac{d\lambda_{iS}}{dt}&= \beta_{iS},\\
\label{eq:rgeff}
\frac{16\pi^2}{\log(10)}\frac{d\lambda_{iS_1S_2}}{dt}&= \beta_{iS_1S_2},
\end{align}
where 
\begin{align}
\label{eq:betasm}
\beta_{h}^{\text{SM}}(\lambda_{ih})&= \frac{27g_1^4}{200}+\frac{9g_1^2g_2^2}{20}+\frac{9g_2^4}{8}-9\big(\frac{g_1^2}{5}+g_2^2\big)\lambda_{ih}+24\lambda_{ih}^2+12\lambda_{ih} Y_t^2-6 Y_t^4 \\
\label{eq:betahfsm}
\beta_{hS}^{\text{SM}}(\lambda_{ihS})&= \frac{{\lambda_{ihS}}}{10}(-9 g_1^2 -45 g_2^2+120\lambda_{ih}+80\lambda_{ihS}+80\lambda_{iS}+60 Y_t^2)
\end{align}
where all the couplings $\lambda_{4h},\lambda_{4hS},\lambda_{4S},\lambda_{4S_1S_2}$ are identified with their corresponding full potential couplings $\lambda_{h},\lambda_{hS},\lambda_{S},\lambda_{S_1S_2}$ and the beta functions $\beta_{4hS}\equiv\beta_{hS},~\beta_{4S}\equiv\beta_{S},~\beta_{4S_1S_2}\equiv\beta_{S_1S_2}$ while $\beta_{4\phi}\equiv\beta_{\phi}\equiv0$. The interaction beta functions $\beta_{h}^{\text{int}}(\lambda_{hS})$, $\beta_{hS}^{\text{int}}(\lambda_{hS},\lambda_{S_1S_2})$ and $\beta_{S,S_1S_2}(\lambda_{hS},\lambda_{S},\lambda_{S_1S_2})$ are given explicitly in each case below.

For $t=\log_{10}{(Q/{\rm GeV})}\sim[16,20]$, where the energy scale $Q$ is in GeV, the beta functions of the Higgs quartic coupling $\lambda_h$ where the full potential of Eq.~(\ref{eq:V4}) is considered are (for $ S,S_1,S_2=\phi,\psi,\chi$)
\begin{align}
\label{eq:betah4}
\beta_{h}&= \beta^{\text{SM}}(\lambda_h)+4(\lambda_{h\phi}^2+\lambda_{h\psi}^2+\lambda_{h\chi}^2), \\
\label{eq:betahf4}
\beta_{hS}&= \beta_{hS}^{\text{SM}}(\lambda_{hS})+4(\lambda_{hS_1} \lambda_{SS_1}+ \lambda_{hS_2} \lambda_{SS_2}),\quad S_1,S_2\neq S,~S_1\neq S_2, \\
\label{eq:betaf4}
\beta_{S}&= 8\lambda_{hS}^2+20\lambda_{S}^2+4(\lambda_{SS_1}^2+\lambda_{SS_2}^2),\quad S_1,S_2\neq S, \\
\label{eq:betaff4}
\beta_{S_1S_2}&= 8(\lambda_{hS_1} \lambda_{hS_2}+\lambda_{S_1S_2} (\lambda_{S_1S_2}+\lambda_{S_1}+\lambda_{S_2}))+4\lambda_{SS_1}\lambda_{SS_2},\quad S_1,S_2\neq S,
\end{align}
where each $\lambda_{\phi\chi}$ must be replaced with $-\lambda_{\phi\chi}$ as mentioned in footnote~\ref{ftnt:lmphchsign}.

Also, the beta functions of the effective 3-field model of Eq.~(\ref{eq:V3}) for $t\sim[12,16]$ (for $S,S_1,S_2=\phi,\chi$) reads as
\begin{align}
\label{eq:betah3}
\beta_{3h}&= \beta_{h}^{\text{SM}}(\lambda_{3h})+4(\lambda_{3h\phi}^2+\lambda_{3h\chi}^2)\\
\label{eq:betahf3}
\beta_{3hS}&= \beta_{hS}^{\text{SM}}(\lambda_{3hS})+4\lambda_{3hS_1}\lambda_{3SS_1},\quad S_1\neq S\\
\label{eq:betaf3}
\beta_{3S}&= 8\lambda_{3hS}^2+20\lambda_{3S}^2+4\lambda_{3SS_1}^2,\quad S_1\neq S \\
\label{eq:betaff3}
\beta_{3S_1S_2}&= 8(\lambda_{3hS_1} \lambda_{3hS_2}+\lambda_{3S_1S_2} (\lambda_{3S_1S_2}+\lambda_{3S_1}+\lambda_{3S_2})),\quad S_1,S_2\neq S 
\end{align}

Finally, the beta function of the effective 2-field model of Eq.~(\ref{eq:V2}) for $t\sim[8,12]$ reads as
\begin{align}
\label{eq:betah2}
\beta_{2h}&= \beta_{h}^{\text{SM}}(\lambda_{2h})+4\lambda_{2h\chi}^2, \\
\label{eq:betahf2}
\beta_{2h\chi}&= \beta_{h\chi}^{\text{SM}}(\lambda_{2h\chi}), \\
\label{eq:betaf2}
\beta_{2\chi}&= 8\lambda_{2h\chi}^2 + 20 \lambda_{2\chi}^2.
\end{align}
Generally, for an $i$-field case with $S_i$'s fields of $(\phi,\psi,\chi)=(S_1,S_2,S_3)$ and $S=S_1,\ldots,S_i$, $i=1,\ldots,\leq3$ reads as
\begin{align}
\label{eq:betah}
\beta_{ih}&= \beta_{h}^{\text{SM}}(\lambda_{ih})+4\sum_{S}\lambda_{ihS}^2, \\
\label{eq:betahf}
\beta_{ihS}&= \beta_{hS}^{\text{SM}}(\lambda_{ihS})+4\sum_{S_1\neq S}{\lambda_{ihS_1} \lambda_{iSS_1}}, \\
\label{eq:betaf}
\beta_{iS}&= 8\lambda_{ihS}^2+20\lambda_{iS}^2+4\sum_{S_1\neq S}\lambda_{iSS_1}^2, \\
\label{eq:betaff}
\beta_{iS_1S_2}&= 8(\lambda_{ihS_1} \lambda_{ihS_2}+\lambda_{iS_1S_2} (\lambda_{iS_1S_2}+\lambda_{iS_1}+\lambda_{iS_2}))+4\lambda_{iSS_1}\lambda_{iSS_2},\quad S_1,S_2\neq S,
\end{align}
where each $\lambda_{\phi\chi}$ must be replaced with $-\lambda_{\phi\chi}$ as mentioned in footnote~\ref{ftnt:lmphchsign}.
\begin{figure}[t]
\begin{center}
\includegraphics[width=0.5\textwidth]{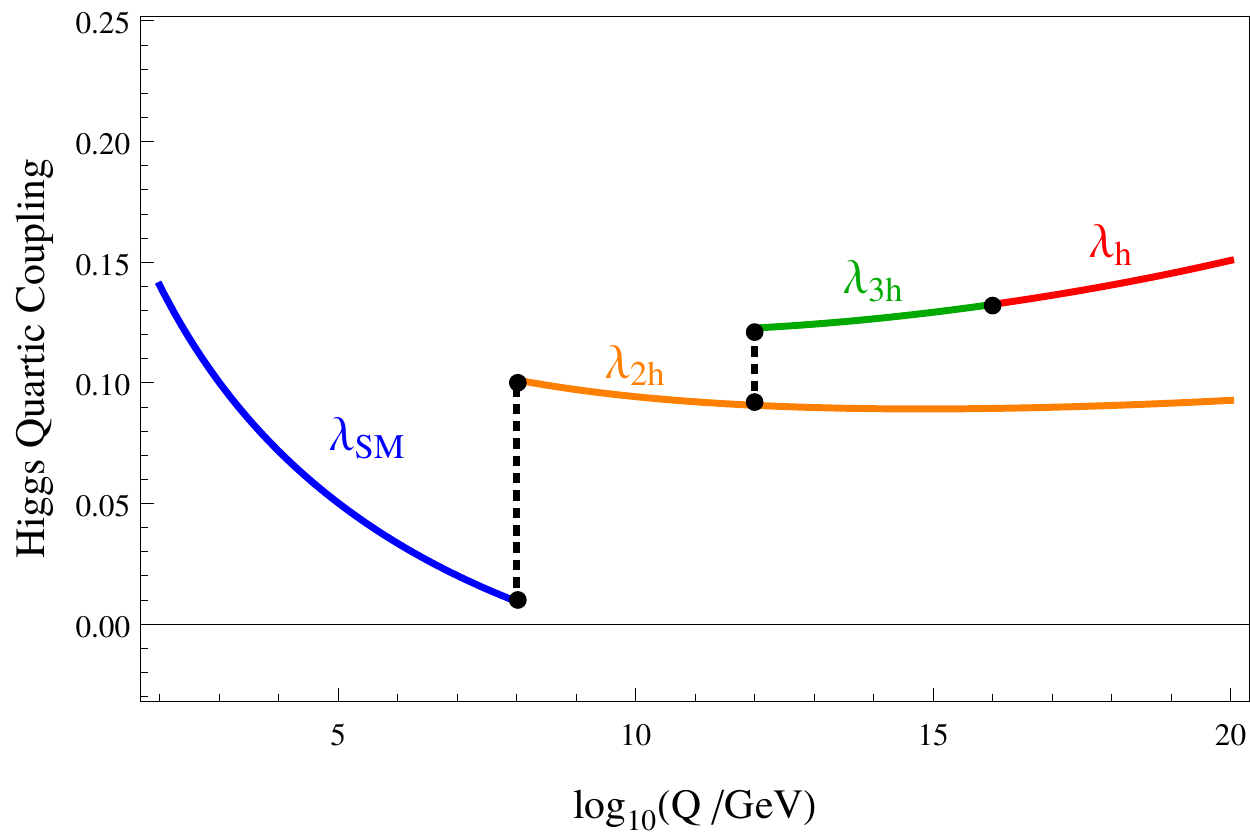}~
\includegraphics[width=0.5\textwidth]{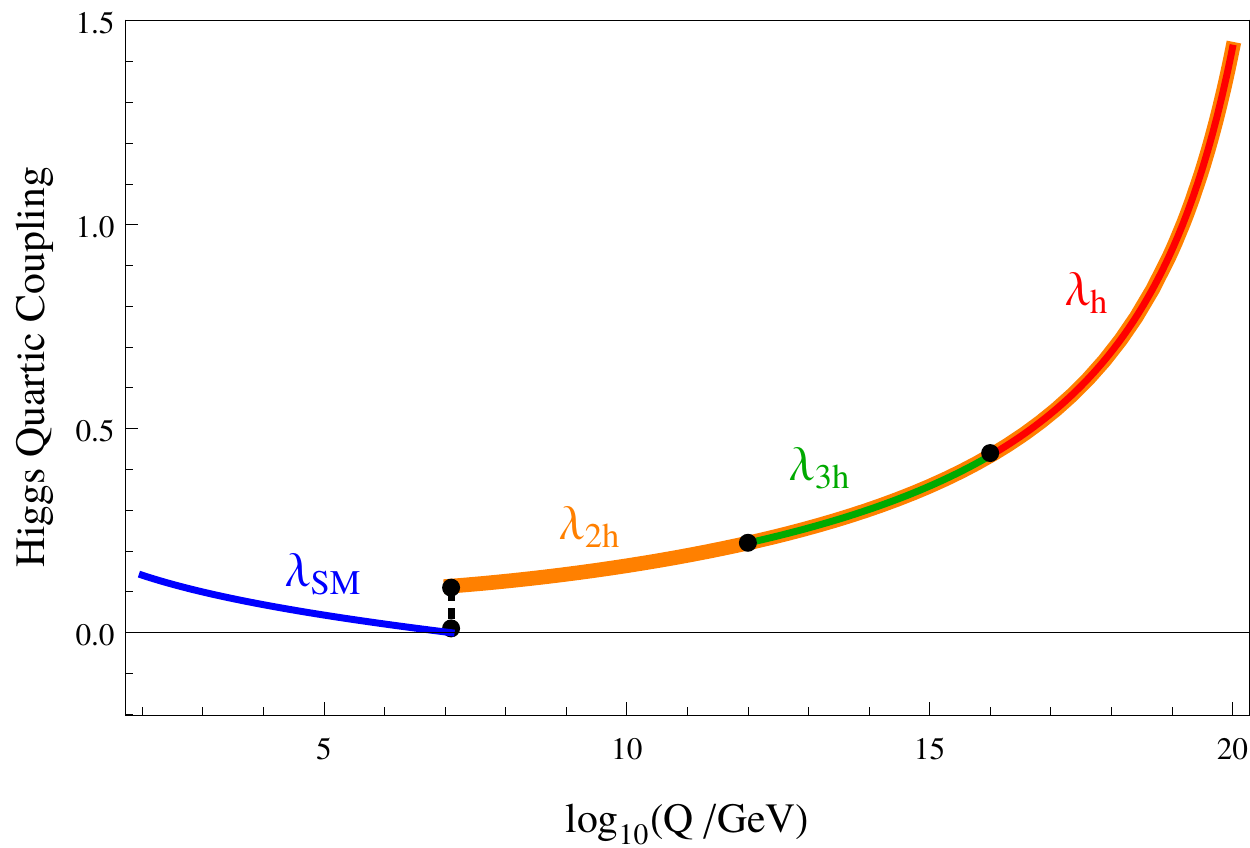}
\caption{\label{fig:EWVS}Running of the Higgs quartic coupling versus the renormalization scale $t\equiv\log_{10}{(Q/{\rm GeV})}$ for the trans-Planckian BP1 of Table~\ref{tab:BP1} and Table~\ref{tab:vsbp3} without/with the inclusion of RHN (left/right). Blue: SM. Orange,~green,~red: SM+MHI at different scales and thresholds. Dotted: thresholds' effects. Similar figures for the sub-Planckian BP2 and the minimal model BP3 can be produced.}
\end{center}
\end{figure} 

The solutions of the RGEs of Eqs.~(\ref{eq:betah4}),~(\ref{eq:betah3})~and~(\ref{eq:betah2}) with the boundary conditions of Eqs.~(\ref{eq:V3mtchngi1}),~(\ref{eq:V3mtchngi2}),~(\ref{eq:V2mtchngi1}),~(\ref{eq:V2mtchngi2}),~and~(\ref{eq:V1mtchngi}) are shown in Fig.~\ref{fig:EWVS} for BP1 presented in Table~\ref{tab:BP1} and Table~\ref{tab:vsbp3}. It is worth mentioning that this BP is shown by the black solid point in the heart of MHI Planck patch in Fig.~\ref{fig:planck}.

We notice from Eq.~(\ref{eq:betasm}) and from the general Eq.~(\ref{eq:betah}) of all cases [Eqs.~(\ref{eq:betah4}), (\ref{eq:betah3}), and (\ref{eq:betah2})] that the coupling of the SM Higgs to the scalars from the beyond SM (BSM) of the MHI helps the SM Higgs potential to continue more stable up to a scale at which the SM is assumed valid. This comes by the addition of the positive contribution $\beta_{h}^{\text{int}}=4\sum_{S}\lambda_{ihS}^2$ to the SM contribution of~(\ref{eq:betasm}) which is dominated by the negative $Y_t$ contribution loops. At a certain threshold of the initial parameter values, this BSM contribution dominates the SM $Y_t$ up to the Planck scale as in Fig.~\ref{fig:EWVS}. Below the parameter threshold, the SM $Y_t$ dominates again making the SM vacuum stable only up to $\mathcal{O}(\text{GUT})$ scale or even earlier depending on the initial jump of the parameters. Moreover, we notice that this behavior is easily achievable at high scales where the contribution comes from the Higgs coupling to more fields and hence the positive contribution in Eq.~(\ref{eq:betah}) will be due to many parameters. Not only this but also the decreasing running of the $Y_t$ to lower values relaxes its negative contribution domination at the EW to the $\Lambda_I$ scale. It is also clear from Eq.~(\ref{eq:betaf}) that the inflation parameters $\lambda_{S}$ run always increasingly in all cases [Eqs.~(\ref{eq:betaf4}), (\ref{eq:betaf3}), and (\ref{eq:betaf2})] in Eq.~(\ref{eq:rgef}) as their beta functions are always positive $\beta_{iS}>0$. Finally, the form of the RGEs Eqs.~(\ref{eq:rgehf}) and (\ref{eq:rgeff}) and the beta functions~(\ref{eq:betahfsm}), the interaction beta function $\beta_{hS}^{\text{int}}=4\sum_{S_1\neq S}{\lambda_{ihS_1} \lambda_{iSS_1}}$ in~(\ref{eq:betahf}) and~(\ref{eq:betaff}) with the negative sign of the $\phi$-$\chi$ interaction term in the MHI potential~(\ref{eq:vmhi}), all alter domination of terms at different scales according to their initial values and the running of the parameters $\lambda_{hS}$ and $\lambda_{S_1S_2}$ in all cases [Eqs.~(\ref{eq:betahf4}), (\ref{eq:betahf3}), (\ref{eq:betahf2}) and Eqs.~(\ref{eq:betaff4}), and (\ref{eq:betaff3})].
\begin{table}[t]
{\fontsize{8}{10}
\begin{center}
\begin{tabular}{cccccccc} 
\hline
\hline
{Par} & $\lambda_{2h\chi}|_{\Lambda_I}$ & $\lambda_{2\chi}|_{\Lambda_I}$ & $\lambda_{3h\phi}|_{\Lambda_{2I}}$ & $\lambda_{3\phi\chi}|_{\Lambda_{2I}}$ & $\lambda_{3\phi}|_{\Lambda_{2I}}$ & $\lambda_{h\psi}|_{\Lambda_{3I}}$ \\\hline
{BP1} & \begin{tabular}{c}$7.45\times10^{-5}$ \\ $7.55\times10^{-5}$ \end{tabular} &\begin{tabular}{c} $6.82\times10^{-8}$\\$3.70 \times 10^{-8}$\end{tabular}  &\begin{tabular}{c} $9.21\times10^{-8}$ \\$3.19\times10^{-9}$ \end{tabular}& \begin{tabular}{c} $-1.30\times10^{-11}$\\$-1.24\times10^{-11}$\end{tabular} & \begin{tabular}{c}$8.00\times10^{-15}$ \\$8.00\times10^{-14}$ \end{tabular} &\begin{tabular}{c} $1.60\times10^{-4}$ \\$1.00\times10^{-6}$ \end{tabular}
\\\hline
{BP2} & \begin{tabular}{c} $5.70\times10^{-2}$\\ $1.83\times10^{-3}$ \end{tabular} &  \begin{tabular}{c} $3.40\times10^{-2}$ \\$1.02\times10^{-2}$ \end{tabular}  &  \begin{tabular}{c} $1.00\times10^{-5}$ \\ $1.07 \times 10^{-2}$ \end{tabular} & \begin{tabular}{c} $-1.60\times10^{-7}$ \\$-3.00\times10^{-7}$ \end{tabular} &\begin{tabular}{c} $7.00\times10^{-5}$ \\$3.16\times10^{-10}$ \end{tabular} & \begin{tabular}{c} $1.60\times10^{-6}$\\$3.16\times10^{-6}$ \end{tabular} \\\hline
\hline
\end{tabular}
\caption{\label{tab:vsbp3}Effective thresholds' free parameters boundary values used in Fig.~\ref{fig:EWVS}.}
\end{center}}
\end{table}

Numerically, in the 2-field model of $V_{2{\rm eff}}(h,\chi)$, the two parameters $\lambda_{2\chi},\lambda_{2h\chi}$ have free initial boundary values at the instability scale $\log_{10}\Lambda_I\sim8$. The Higgs quartic coupling $\lambda_{2h}$ boundary value at $\Lambda_I$ is matched to the SM Higgs quartic coupling according to the inverted relation of Eq.~(\ref{eq:V1mtchngi}) as 
\begin{align}
\label{eq:V1mtchngiinv}
\lambda_{2h}\big|_{\Lambda_I}=\Big[\lambda_{\text{SM}}+\frac{\lambda^2_{2h\chi}}{\lambda_{2\chi}}\Big]\Big|_{\Lambda_I}\approx\frac{\lambda^2_{2h\chi}}{\lambda_{2\chi}}\Big|_{\Lambda_I}.
\end{align}
where $\lambda_{\text{SM}}(\Lambda_I)\approx0$ and $\lambda_{2h}\big|_{\Lambda_I}$ represents the first coupling jump at $\Lambda_I$ in Fig.~\ref{fig:EWVS} as determined by the values given in Table~\ref{tab:vsbp3}. Similarly, from Eq.~(\ref{eq:V2mtchngi1})
\begin{align}
\label{eq:V2mtchngiinv}
\lambda_{3h}\big|_{\Lambda_{2I}}=\Big[\lambda_{2h}+\frac{\lambda^2_{3h\phi}}{\lambda_{3\phi}}\Big]\Big|_{\Lambda_{2I}}
\end{align}
where $\log_{10}\Lambda_{2I}\sim12$ is the instability scale of the effective 2-field model $V_{2{\rm eff}}(h,\chi)$ and the fraction in Eq.~(\ref{eq:V2mtchngiinv}) represents the second coupling jump in Fig.~\ref{fig:EWVS}. Finally, the third jump at the instability scale $\log_{10}\Lambda_{3I}\sim16$ of the 3-field $V_{3{\rm eff}}(h,\phi,\chi)$ model is determined by Eq.~(\ref{eq:V3mtchngi1}) as follows
\begin{align}
\label{eq:V2mtchngiinv}
\lambda_{h}\big|_{\Lambda_{3I}}=\Big[\lambda_{3h}+\frac{\lambda^2_{3h\psi}}{\lambda_{3\psi}}\Big]\Big|_{\Lambda_{3I}}.
\end{align}
Nevertheless, we see that the $\lambda_{2h}$ coupling contribution stabilizes the EW Higgs vacuum up to the GUT scale and even beyond, and that the behavior is safe as the contribution from $\lambda_{3h}$ and $\lambda_{h}$ improve the situation. Suitable numerical values of the free parameters can be set so that $\lambda_{3h,2h,h}$ coincide at high scales so that the left figure of Fig.~\ref{fig:EWVS} appears in the right one.

The contribution of the RHN coupling $Y_{\nu}$ of~(\ref{eq:rhnlag}) modifies the SM top-Yukawa coupling $Y_{t}$ and Higgs quartic coupling $\lambda_h$ one-loop $\beta_t$- and $\beta_h$-functions by $\beta_{t\nu}=Y_{t}Y_{\nu}^2$ and $\beta_{h\nu}=2Y_{\nu}^2(4\lambda_h-Y_{\nu}^2)$, respectively. Also the $Y_{\nu}$ RGE is
\begin{align}\label{eq:rhnrge}
\frac{16\pi^2}{\log(10)} \dfrac{dY_{\nu}}{dt}&= \beta_\nu=Y_{\nu}\Big(\frac{5}{2}Y_{\nu}^2+3Y_{t}^2-\frac{9}{4}g_2^2-\frac{3}{4}g_1^2\Big).
\end{align}
As mentioned in Sec.~\ref{sec:reheat}, the contribution of the $Y_{\nu}$ ($\beta_{t\nu}>0$ and $\beta_{H\nu}<0$) in the instability of the SM Higgs vacuum are negligible, compared to their SM counterparts, as long as $Y_{\nu}\lsim0.3$, as dictated by the 	sub-Planckian case of inflation. Moreover, if $Y_{\nu}\lsim0.6$, this would bring the SM instability scale $\Lambda_I$ to about $\mathcal{O}(10^6)~{\rm GeV}$, and the stability of the potential is insured by the Higgs couplings to the inflation sector fields as in Fig.~\ref{fig:EWVS}~(right panel). In either case, the $Y_{\nu}$ RGE~(\ref{eq:rhnrge}) clarifies that $Y_{\nu}$ is increasing and contributions from the $V_{3\rm eff}$ model~(\ref{eq:V3}) at the $\Lambda_{2I}$ scale, and maybe the full $V$ model~(\ref{eq:V4}) at the $\Lambda_{3I}$ scale, should be included as in Fig.~\ref{fig:EWVS}~(right panel) to ensure the vacuum stability up to the Planck scale. For $Y_{\nu}\lsim0.6$, a lower limit is set from Eq.~(\ref{eq:mnu}) on the RHN mass is set $M_N\gsim~Y_{\nu}^2v^2/m_{\nu}\sim10^{10}~{\rm GeV}$.

At this point, we mention the effect of the SM Higgs mixing with the MHI scalars on the SM Higgs mass and interactions. As a very good approximation, we consider the effective 2-field model $V_{2{\rm eff}}(h,\chi)$. In this case, the $2\times2$ mass matrix of $h$ and $\chi$ is
\begin{equation}
{\cal M}_{h\chi}^2=2
\begin{pmatrix}
\lambda_{2h}v^2 & \lambda_{2h\chi}vv_\chi \\
\lambda_{2h\chi}vv_\chi & \lambda_{2\chi}v_\chi^2
\end{pmatrix}.
\end{equation}
with exact squared masses
\begin{equation}
m^2_{h,\chi}=\lambda_{2h}v^2+\lambda_{2\chi}v_\chi^2\mp
\sqrt{(\lambda_{2h}v^2-\lambda_{2\chi}v_\chi^2)^2+4\lambda_{2h\chi}^2 v^2 v_{\chi}^2}.
\end{equation}
and the couplings were substituted at the EW scale such that $\lambda_{2h\chi}|_{\rm EW}\sim5.79\times10^{-5}$ and $\lambda_{2\chi}|_{\rm EW}\sim6.51\times10^{-8}$ consistent with Table~\ref{tab:vsbp3}, where the EW scale at the top-quark mass $m_t=172.69~{\rm GeV}$. As discussed in Ref.~\cite{Elias-Miro:2012eoi}, for $m_\psi^2\gsim m_\phi^2\gg\lambda_{2\chi}v_\chi^2\gg\lambda_{2h}v^2$, the masses squared eigenvalues are expanded to the first order of $v^2/v_\chi^2$ and the `seesaw-like' corrected masses at the EW scale are
\begin{align}
m_h^2 & \approx2v^2\Big[\lambda_{2h}-\frac{\lambda_{2h\chi}^2}{\lambda_{2\chi}}\Big]\Big|_{{\rm EW}}\\
m_\chi^2 & \approx2 v_\chi^2\Big[\lambda_{2\chi}+\frac{\lambda_{2h\chi}^2}{\lambda_{2\chi}}\frac{v^2}{v_\chi^2}\Big]\Big|_{{\rm EW}}
\end{align}
Accordingly, for the SM Higgs mass $m_h=125.25~{\rm GeV}$, we have the following boundary constraint for the SM effective Higgs quartic coupling
\begin{align}\label{eq:lamsmeff}
\lambda_{\rm eff}=\Big[\lambda_{2h}-\frac{\lambda_{2h\chi}^2}{\lambda_\chi}\Big]\Big|_{{\rm EW}}\sim0.12
\end{align}
That is, the SM MC relation (\ref{eq:V1mtchngiinv}) holds at the EW scale too for the BPs in Table~\ref{tab:vsbp3}: for BP1 with $(\lambda_{2h\chi},\lambda_{2\chi})\big|_{\rm EW}=(5.68\times10^{-5},6.50\times10^{-8})$, while for fixing $m_h$, $\lambda_{2h}\big|_{\rm EW}\sim1.79\times10^{-1}$ consistent with (\ref{eq:lamsmeff}). Also, for BP2 with $(\lambda_{2h\chi},\lambda_{2\chi})\big|_{\rm EW}=(4.07\times10^{-2},3.05\times10^{-2})$ and for fixing $m_h$, $\lambda_{2h}\big|_{\rm EW}\sim1.84\times10^{-1}$ consistent with (\ref{eq:lamsmeff}).
Also, the mixing angle
\begin{equation}\label{eq:hcmixsmvs2ex}
\tan2\theta_{h\chi}
=\frac{2\lambda_{2h\chi}vv_\chi}{\lambda_{2\chi}v_\chi^2-\lambda_{2h}v^2}\Big|_{\rm EW}
\approx\frac{2\lambda_{2h\chi}}{\lambda_{2\chi}}\frac{v}{v_\chi}
\Big(1+\frac{\lambda_{2h}}{\lambda_{2\chi}}\big(\frac{v}{v_\chi}\big)^2\Big)\Big|_{\rm EW}
\sim\mathcal{O}(10^{-7}),
\end{equation}
and this preserves the SM Higgs physics up to the Planck scale for the BPs in Table~\ref{tab:BP1} and Table~\ref{tab:vsbp3} as checked for the running of the mixing angle (\ref{eq:hcmixsmvs2ex}).

Now we discuss Higgs quantum fluctuations during inflation, the EW vacuum may be destabilized during inflation by quantum fluctuations of the Higgs field if $H^{\rm inf} > \Lambda_I$ \cite{Espinosa:2007qp,Lebedev:2012sy,Kamada:2014ufa,Ema:2017ckf,Lebedev:2021xey}. In fact, our model is safe against this kind of destabilization, since the effective Higgs mass during inflation $m_h(\phi) \gg H^{\rm inf}(\phi)$ as indicated in Fig. \ref{fig:mhub} in both cases of large and small field inflation. In that case, the quantum fluctuations are suppressed. In addition, quantum fluctuations may destabilize the EW vacuum after inflation if the Higgs inflaton coupling $> 3\times 10^{-8}$ for typical Higgs quartic coupling $\lambda_h =-10^{-2}$ as indicated in Ref.\cite{Lebedev:2021xey}. However, in our model Higgs quartic coupling $\lambda_h$ is positive up to the Planck scale. Therefore, again we expect that our model is safe against quantum fluctuations after inflation ends.

A final comment is in order. As the Higgs interactions with the other inflation fields is exploited to maintain the Higgs vacuum stability, conversely, the effect of the running couplings on the inflation observables is also considered. The running couplings initial values were taken such that they give rise to consistent inflation observables, and thus ensured the validity of the observables values.

\section{\label{sec:conclusions}CONCLUSIONS}
We studied the connection of a modified hybrid inflation model to the standard model electroweak vacuum stability. We have extended the inflation sector with an extra scalar $\chi$, which is, as well as $\phi$ and $\psi$, a SM singlet. The scalar $\chi$ may be an extra $U(1)$ Higgs field and $\psi$ can be assigned as a GUT gauge group Higgs while the inflaton $\phi$ is singlet under all gauge groups. The complete discussion of a more general model with a specific gauge group extending the SM one, and its phenomenology will be provided in a future work. 

The modification results in an inflation potential in which the inflaton rolls down near a hilltop in the valley of the other hybrid fields which are stabilized at their false vacua during the inflation. The usual hybrid inflation problem of large spectral tilt is then resolved since the inflation occurs mostly in the negative curvature part of the potential, before the inflection point. 

The parameter space was analyzed and the inflation observables were calculated in both trans-Planckian and sub-Planckian cases in consistency with the recent Planck/BICEP observations. We have provided the couplings of the SM Higgs with the inflation singlet in order to stabilize the electroweak vacuum up to the Planck scale. We have studied the decays of the inflation to right-handed neutrinos, that allow for reheating the universe. Moreover, quantum corrections to the inflation potential were taken into account. We found that an upper bound on the reheating temperature will suppress the contributions of quantum corrections to inflation effective potential. We found that the neutrino Yukawa coupling can be of order ${\cal O}(1)$, which reduces the instability scale of the EW vacuum. We have shown that the threshold corrections will stabilize the EW vacuum even in this case. 

\section*{ACKNOWLEDGMENTS}
The authors would like to thank Qaisar Shafi for the useful discussions. 
The work of M.~I., M.~A. and A.~M. is partially supported by the Science, Technology~$\&$~Innovation Funding Authority~(STDF) under Grant No.~33495.

\bibliographystyle{JHEP}\bibliography{Bib}
\end{document}